\documentclass[longauth]{aa}  

\usepackage{graphicx}
\usepackage{txfonts}
\usepackage{tabularx}
\usepackage{multirow}
\usepackage{booktabs}
\usepackage{float}
\usepackage{xcolor}
\usepackage{natbib}
\usepackage{comment}
\usepackage[version=4]{mhchem}
\bibpunct{(}{)}{;}{a}{}{,} 
\usepackage[breaklinks, colorlinks, citecolor=blue, linkcolor=blue]{hyperref}
\makeatletter
\renewcommand*\aa@pageof{, page \thepage{} of \pageref*{LastPage}}
\makeatother

\newlength{\bibitemsep}
\newlength{\bibparskip}
\let\oldthebibliography\thebibliography
\renewcommand\thebibliography[1]{%
  \oldthebibliography{#1}%
  \setlength{\parskip}{\bibitemsep}%
  \setlength{\itemsep}{\bibparskip}%
}

\begin{document} 

\title{Searching for Hot Water World Candidates with CHEOPS\thanks{This article uses data from the CHEOPS Guaranteed Time Observation programme CH\_PR140068. The raw and detrended photometric time-series data are available in electronic form at the CDS via anonymous ftp to cdsarc.ustrasbg.fr (130.79.128.5) or via \url{https://cdsarc.cds.unistra.fr/viz-bin/cat/J/A+A/XXX/YYY}}}
\subtitle{Refining the radii and analysing the internal structures and atmospheric lifetimes of TOI-238~b and TOI-1685~b}

\author{
    J.~A.~Egger\inst{\ref{inst:1}}\,$^{\href{https://orcid.org/0000-0003-1628-4231}{\protect\includegraphics[height=0.19cm]{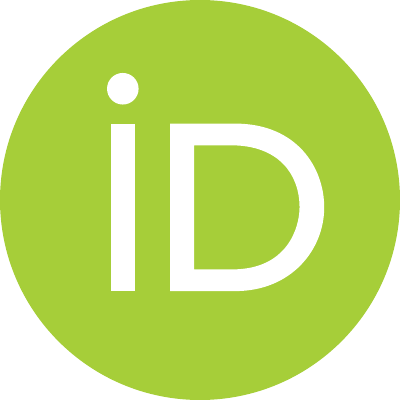}}}$, 
    D.~Kubyshkina\inst{\ref{inst:2}}, 
    Y.~Alibert\inst{\ref{inst:3},\ref{inst:1}}\,$^{\href{https://orcid.org/0000-0002-4644-8818}{\protect\includegraphics[height=0.19cm]{figures/orcid.pdf}}}$, 
    H.~P.~Osborn\inst{\ref{inst:3},\ref{inst:4}}\,$^{\href{https://orcid.org/0000-0002-4047-472}{\protect\includegraphics[height=0.19cm]{figures/orcid.pdf}}}$, 
    A.~Bonfanti\inst{\ref{inst:2}}\,$^{\href{https://orcid.org/0000-0002-1916-5935}{\protect\includegraphics[height=0.19cm]{figures/orcid.pdf}}}$, 
    T.~G.~Wilson\inst{\ref{inst:5}}\,$^{\href{https://orcid.org/0000-0001-8749-1962}{\protect\includegraphics[height=0.19cm]{figures/orcid.pdf}}}$, 
    A.~Brandeker\inst{\ref{inst:6}}\,$^{\href{https://orcid.org/0000-0002-7201-7536}{\protect\includegraphics[height=0.19cm]{figures/orcid.pdf}}}$, 
    M.~N.~Günther\inst{\ref{inst:7}}\,$^{\href{https://orcid.org/0000-0002-3164-9086}{\protect\includegraphics[height=0.19cm]{figures/orcid.pdf}}}$, 
    M.~Lendl\inst{\ref{inst:8}}\,$^{\href{https://orcid.org/0000-0001-9699-1459}{\protect\includegraphics[height=0.19cm]{figures/orcid.pdf}}}$, 
    D.~Kitzmann\inst{\ref{inst:1},\ref{inst:3}}, 
    L.~Fossati\inst{\ref{inst:2}}\,$^{\href{https://orcid.org/0000-0003-4426-9530}{\protect\includegraphics[height=0.19cm]{figures/orcid.pdf}}}$, 
    C.~Mordasini\inst{\ref{inst:1},\ref{inst:3}}, 
    S.~G.~Sousa\inst{\ref{inst:9}}\,$^{\href{https://orcid.org/0000-0001-9047-2965}{\protect\includegraphics[height=0.19cm]{figures/orcid.pdf}}}$, 
    V.~Adibekyan\inst{\ref{inst:9}}\,$^{\href{https://orcid.org/0000-0002-0601-6199}{\protect\includegraphics[height=0.19cm]{figures/orcid.pdf}}}$, 
    M.~Fridlund\inst{\ref{inst:10},\ref{inst:11}}\,$^{\href{https://orcid.org/0000-0002-0855-8426}{\protect\includegraphics[height=0.19cm]{figures/orcid.pdf}}}$, 
    C.~Pezzotti\inst{\ref{inst:12}}, 
    D.~Gandolfi\inst{\ref{inst:13}}\,$^{\href{https://orcid.org/0000-0001-8627-9628}{\protect\includegraphics[height=0.19cm]{figures/orcid.pdf}}}$, 
    S.~Ulmer-Moll\inst{\ref{inst:8},\ref{inst:1},\ref{inst:12}}\,$^{\href{https://orcid.org/0000-0003-2417-7006}{\protect\includegraphics[height=0.19cm]{figures/orcid.pdf}}}$, 
    R.~Alonso\inst{\ref{inst:14},\ref{inst:15}}\,$^{\href{https://orcid.org/0000-0001-8462-8126}{\protect\includegraphics[height=0.19cm]{figures/orcid.pdf}}}$, 
    T.~Bárczy\inst{\ref{inst:16}}\,$^{\href{https://orcid.org/0000-0002-7822-4413}{\protect\includegraphics[height=0.19cm]{figures/orcid.pdf}}}$, 
    D.~Barrado~Navascues\inst{\ref{inst:17}}\,$^{\href{https://orcid.org/0000-0002-5971-9242}{\protect\includegraphics[height=0.19cm]{figures/orcid.pdf}}}$, 
    S.~C.~C.~Barros\inst{\ref{inst:9},\ref{inst:18}}\,$^{\href{https://orcid.org/0000-0003-2434-3625}{\protect\includegraphics[height=0.19cm]{figures/orcid.pdf}}}$, 
    W.~Baumjohann\inst{\ref{inst:2}}\,$^{\href{https://orcid.org/0000-0001-6271-0110}{\protect\includegraphics[height=0.19cm]{figures/orcid.pdf}}}$, 
    W.~Benz\inst{\ref{inst:1},\ref{inst:3}}\,$^{\href{https://orcid.org/0000-0001-7896-6479}{\protect\includegraphics[height=0.19cm]{figures/orcid.pdf}}}$, 
    N.~Billot\inst{\ref{inst:8}}\,$^{\href{https://orcid.org/0000-0003-3429-3836}{\protect\includegraphics[height=0.19cm]{figures/orcid.pdf}}}$, 
    L.~Borsato\inst{\ref{inst:19}}\,$^{\href{https://orcid.org/0000-0003-0066-9268}{\protect\includegraphics[height=0.19cm]{figures/orcid.pdf}}}$, 
    C.~Broeg\inst{\ref{inst:1},\ref{inst:3}}\,$^{\href{https://orcid.org/0000-0001-5132-2614}{\protect\includegraphics[height=0.19cm]{figures/orcid.pdf}}}$, 
    A.~Collier~Cameron\inst{\ref{inst:20}}\,$^{\href{https://orcid.org/0000-0002-8863-7828}{\protect\includegraphics[height=0.19cm]{figures/orcid.pdf}}}$, 
    A.~C.~M.~Correia\inst{\ref{inst:21}}\,$^{\href{https://orcid.org/0000-0002-8946-8579}{\protect\includegraphics[height=0.19cm]{figures/orcid.pdf}}}$, 
    Sz.~Csizmadia\inst{\ref{inst:22}}\,$^{\href{https://orcid.org/0000-0001-6803-9698}{\protect\includegraphics[height=0.19cm]{figures/orcid.pdf}}}$, 
    P.~E.~Cubillos\inst{\ref{inst:2},\ref{inst:23}}, 
    M.~B.~Davies\inst{\ref{inst:24}}\,$^{\href{https://orcid.org/0000-0001-6080-1190}{\protect\includegraphics[height=0.19cm]{figures/orcid.pdf}}}$, 
    M.~Deleuil\inst{\ref{inst:25}}\,$^{\href{https://orcid.org/0000-0001-6036-0225}{\protect\includegraphics[height=0.19cm]{figures/orcid.pdf}}}$, 
    A.~Deline\inst{\ref{inst:8}}, 
    O.~D.~S.~Demangeon\inst{\ref{inst:9},\ref{inst:18}}\,$^{\href{https://orcid.org/0000-0001-7918-0355}{\protect\includegraphics[height=0.19cm]{figures/orcid.pdf}}}$, 
    B.-O.~Demory\inst{\ref{inst:3},\ref{inst:1}}\,$^{\href{https://orcid.org/0000-0002-9355-5165}{\protect\includegraphics[height=0.19cm]{figures/orcid.pdf}}}$, 
    A.~Derekas\inst{\ref{inst:26}}, 
    B.~Edwards\inst{\ref{inst:27}}, 
    D.~Ehrenreich\inst{\ref{inst:8},\ref{inst:28}}\,$^{\href{https://orcid.org/0000-0001-9704-5405}{\protect\includegraphics[height=0.19cm]{figures/orcid.pdf}}}$, 
    A.~Erikson\inst{\ref{inst:22}}, 
    A.~Fortier\inst{\ref{inst:1},\ref{inst:3}}\,$^{\href{https://orcid.org/0000-0001-8450-3374}{\protect\includegraphics[height=0.19cm]{figures/orcid.pdf}}}$, 
    K.~Gazeas\inst{\ref{inst:29}}\,$^{\href{https://orcid.org/0000-0002-8855-3923}{\protect\includegraphics[height=0.19cm]{figures/orcid.pdf}}}$, 
    M.~Gillon\inst{\ref{inst:30}}\,$^{\href{https://orcid.org/0000-0003-1462-7739}{\protect\includegraphics[height=0.19cm]{figures/orcid.pdf}}}$, 
    M.~Güdel\inst{\ref{inst:31}}, 
    A.~Heitzmann\inst{\ref{inst:8}}\,$^{\href{https://orcid.org/0000-0002-8091-7526}{\protect\includegraphics[height=0.19cm]{figures/orcid.pdf}}}$, 
    Ch.~Helling\inst{\ref{inst:2},\ref{inst:32}}, 
    K.~G.~Isaak\inst{\ref{inst:7}}\,$^{\href{https://orcid.org/0000-0001-8585-1717}{\protect\includegraphics[height=0.19cm]{figures/orcid.pdf}}}$, 
    L.~L.~Kiss\inst{\ref{inst:33},\ref{inst:34}}, 
    J.~Korth\inst{\ref{inst:35}}\,$^{\href{https://orcid.org/0000-0002-0076-6239}{\protect\includegraphics[height=0.19cm]{figures/orcid.pdf}}}$, 
    K.~W.~F.~Lam\inst{\ref{inst:22}}\,$^{\href{https://orcid.org/0000-0002-9910-6088}{\protect\includegraphics[height=0.19cm]{figures/orcid.pdf}}}$, 
    J.~Laskar\inst{\ref{inst:36}}\,$^{\href{https://orcid.org/0000-0003-2634-789X}{\protect\includegraphics[height=0.19cm]{figures/orcid.pdf}}}$, 
    A.~Lecavelier~des~Etangs\inst{\ref{inst:37}}\,$^{\href{https://orcid.org/0000-0002-5637-5253}{\protect\includegraphics[height=0.19cm]{figures/orcid.pdf}}}$, 
    A.~Luntzer\inst{\ref{inst:31}}, 
    R.~Luque\inst{\ref{inst:38},\ref{inst:39}}\,$^{\href{https://orcid.org/0000-0002-4671-2957}{\protect\includegraphics[height=0.19cm]{figures/orcid.pdf}}}$, 
    D.~Magrin\inst{\ref{inst:19}}\,$^{\href{https://orcid.org/0000-0003-0312-313X}{\protect\includegraphics[height=0.19cm]{figures/orcid.pdf}}}$, 
    P.~F.~L.~Maxted\inst{\ref{inst:40}}\,$^{\href{https://orcid.org/0000-0003-3794-1317}{\protect\includegraphics[height=0.19cm]{figures/orcid.pdf}}}$, 
    B.~Merín\inst{\ref{inst:41}}\,$^{\href{https://orcid.org/0000-0002-8555-3012}{\protect\includegraphics[height=0.19cm]{figures/orcid.pdf}}}$, 
    M.~Munari\inst{\ref{inst:42}}\,$^{\href{https://orcid.org/0000-0003-0990-050X}{\protect\includegraphics[height=0.19cm]{figures/orcid.pdf}}}$, 
    V.~Nascimbeni\inst{\ref{inst:19}}\,$^{\href{https://orcid.org/0000-0001-9770-1214}{\protect\includegraphics[height=0.19cm]{figures/orcid.pdf}}}$, 
    G.~Olofsson\inst{\ref{inst:6}}\,$^{\href{https://orcid.org/0000-0003-3747-7120}{\protect\includegraphics[height=0.19cm]{figures/orcid.pdf}}}$, 
    R.~Ottensamer\inst{\ref{inst:31}}, 
    I.~Pagano\inst{\ref{inst:42}}\,$^{\href{https://orcid.org/0000-0001-9573-4928}{\protect\includegraphics[height=0.19cm]{figures/orcid.pdf}}}$, 
    E.~Pallé\inst{\ref{inst:14},\ref{inst:15}}\,$^{\href{https://orcid.org/0000-0003-0987-1593}{\protect\includegraphics[height=0.19cm]{figures/orcid.pdf}}}$, 
    G.~Peter\inst{\ref{inst:43}}\,$^{\href{https://orcid.org/0000-0001-6101-2513}{\protect\includegraphics[height=0.19cm]{figures/orcid.pdf}}}$, 
    D.~Piazza\inst{\ref{inst:44}}, 
    G.~Piotto\inst{\ref{inst:19},\ref{inst:45}}\,$^{\href{https://orcid.org/0000-0002-9937-6387}{\protect\includegraphics[height=0.19cm]{figures/orcid.pdf}}}$, 
    D.~Pollacco\inst{\ref{inst:5}}, 
    D.~Queloz\inst{\ref{inst:4},\ref{inst:46}}\,$^{\href{https://orcid.org/0000-0002-3012-0316}{\protect\includegraphics[height=0.19cm]{figures/orcid.pdf}}}$, 
    R.~Ragazzoni\inst{\ref{inst:19},\ref{inst:45}}\,$^{\href{https://orcid.org/0000-0002-7697-5555}{\protect\includegraphics[height=0.19cm]{figures/orcid.pdf}}}$, 
    N.~Rando\inst{\ref{inst:7}}, 
    H.~Rauer\inst{\ref{inst:22},\ref{inst:47}}\,$^{\href{https://orcid.org/0000-0002-6510-1828}{\protect\includegraphics[height=0.19cm]{figures/orcid.pdf}}}$, 
    I.~Ribas\inst{\ref{inst:48},\ref{inst:49}}\,$^{\href{https://orcid.org/0000-0002-6689-0312}{\protect\includegraphics[height=0.19cm]{figures/orcid.pdf}}}$, 
    N.~C.~Santos\inst{\ref{inst:9},\ref{inst:18}}\,$^{\href{https://orcid.org/0000-0003-4422-2919}{\protect\includegraphics[height=0.19cm]{figures/orcid.pdf}}}$, 
    G.~Scandariato\inst{\ref{inst:42}}\,$^{\href{https://orcid.org/0000-0003-2029-0626}{\protect\includegraphics[height=0.19cm]{figures/orcid.pdf}}}$, 
    D.~Ségransan\inst{\ref{inst:8}}\,$^{\href{https://orcid.org/0000-0003-2355-8034}{\protect\includegraphics[height=0.19cm]{figures/orcid.pdf}}}$, 
    A.~E.~Simon\inst{\ref{inst:1},\ref{inst:3}}\,$^{\href{https://orcid.org/0000-0001-9773-2600}{\protect\includegraphics[height=0.19cm]{figures/orcid.pdf}}}$, 
    A.~M.~S.~Smith\inst{\ref{inst:22}}\,$^{\href{https://orcid.org/0000-0002-2386-4341}{\protect\includegraphics[height=0.19cm]{figures/orcid.pdf}}}$, 
    R.~Southworth\inst{\ref{inst:50}}, 
    M.~Stalport\inst{\ref{inst:12},\ref{inst:30}}, 
    S.~Sulis\inst{\ref{inst:25}}\,$^{\href{https://orcid.org/0000-0001-8783-526X}{\protect\includegraphics[height=0.19cm]{figures/orcid.pdf}}}$, 
    Gy.~M.~Szabó\inst{\ref{inst:26},\ref{inst:51}}\,$^{\href{https://orcid.org/0000-0002-0606-7930}{\protect\includegraphics[height=0.19cm]{figures/orcid.pdf}}}$, 
    S.~Udry\inst{\ref{inst:8}}\,$^{\href{https://orcid.org/0000-0001-7576-6236}{\protect\includegraphics[height=0.19cm]{figures/orcid.pdf}}}$, 
    V.~Van~Grootel\inst{\ref{inst:12}}\,$^{\href{https://orcid.org/0000-0003-2144-4316}{\protect\includegraphics[height=0.19cm]{figures/orcid.pdf}}}$, 
    J.~Venturini\inst{\ref{inst:8}}\,$^{\href{https://orcid.org/0000-0001-9527-2903}{\protect\includegraphics[height=0.19cm]{figures/orcid.pdf}}}$, 
    E.~Villaver\inst{\ref{inst:14},\ref{inst:15}}, 
    N.~A.~Walton\inst{\ref{inst:52}}\,$^{\href{https://orcid.org/0000-0003-3983-8778}{\protect\includegraphics[height=0.19cm]{figures/orcid.pdf}}}$, 
    S.~Wolf\inst{\ref{inst:1}}, and
    D.~Wolter\inst{\ref{inst:22}}
}

\authorrunning{J. A. Egger et al.}
   
\institute{
    \label{inst:1} Space Research and Planetary Sciences, Physics Institute, University of Bern, Gesellschaftsstrasse 6, 3012 Bern, Switzerland \and
    \label{inst:2} Space Research Institute, Austrian Academy of Sciences, Schmiedlstrasse 6, A-8042 Graz, Austria \and
    \label{inst:3} Center for Space and Habitability, University of Bern, Gesellschaftsstrasse 6, 3012 Bern, Switzerland \and
    \label{inst:4} ETH Zurich, Department of Physics, Wolfgang-Pauli-Strasse 2, CH-8093 Zurich, Switzerland \and
    \label{inst:5} Department of Physics, University of Warwick, Gibbet Hill Road, Coventry CV4 7AL, United Kingdom \and
    \label{inst:6} Department of Astronomy, Stockholm University, AlbaNova University Center, 10691 Stockholm, Sweden \and
    \label{inst:7} European Space Agency (ESA), European Space Research and Technology Centre (ESTEC), Keplerlaan 1, 2201 AZ Noordwijk, The Netherlands \and
    \label{inst:8} Observatoire astronomique de l'Université de Genève, Chemin Pegasi 51, 1290 Versoix, Switzerland \and
    \label{inst:9} Instituto de Astrofisica e Ciencias do Espaco, Universidade do Porto, CAUP, Rua das Estrelas, 4150-762 Porto, Portugal \and
    \label{inst:10} Leiden Observatory, University of Leiden, PO Box 9513, 2300 RA Leiden, The Netherlands \and
    \label{inst:11} Department of Space, Earth and Environment, Chalmers University of Technology, Onsala Space Observatory, 439 92 Onsala, Sweden \and
    \label{inst:12} Space sciences, Technologies and Astrophysics Research (STAR) Institute, Université de Liège, Allée du 6 Août 19C, 4000 Liège, Belgium \and
    \label{inst:13} Dipartimento di Fisica, Università degli Studi di Torino, via Pietro Giuria 1, I-10125, Torino, Italy \and
    \label{inst:14} Instituto de Astrofísica de Canarias, Vía Láctea s/n, 38200 La Laguna, Tenerife, Spain \and
    \label{inst:15} Departamento de Astrofísica, Universidad de La Laguna, Astrofísico Francisco Sanchez s/n, 38206 La Laguna, Tenerife, Spain \and
    \label{inst:16} Admatis, 5. Kandó Kálmán Street, 3534 Miskolc, Hungary \and
    \label{inst:17} Depto. de Astrofísica, Centro de Astrobiología (CSIC-INTA), ESAC campus, 28692 Villanueva de la Cañada (Madrid), Spain \and
    \label{inst:18} Departamento de Fisica e Astronomia, Faculdade de Ciencias, Universidade do Porto, Rua do Campo Alegre, 4169-007 Porto, Portugal \and
    \label{inst:19} INAF, Osservatorio Astronomico di Padova, Vicolo dell'Osservatorio 5, 35122 Padova, Italy \and
    \label{inst:20} Centre for Exoplanet Science, SUPA School of Physics and Astronomy, University of St Andrews, North Haugh, St Andrews KY16 9SS, UK \and
    \label{inst:21} CFisUC, Departamento de Física, Universidade de Coimbra, 3004-516 Coimbra, Portugal \and
    \label{inst:22} Institute of Planetary Research, German Aerospace Center (DLR), Rutherfordstrasse 2, 12489 Berlin, Germany \and
    \label{inst:23} INAF, Osservatorio Astrofisico di Torino, Via Osservatorio, 20, I-10025 Pino Torinese To, Italy \and
    \label{inst:24} Centre for Mathematical Sciences, Lund University, Box 118, 221 00 Lund, Sweden \and
    \label{inst:25} Aix Marseille Univ, CNRS, CNES, LAM, 38 rue Frédéric Joliot-Curie, 13388 Marseille, France \and
    \label{inst:26} ELTE Gothard Astrophysical Observatory, 9700 Szombathely, Szent Imre h. u. 112, Hungary \and
    \label{inst:27} SRON Netherlands Institute for Space Research, Niels Bohrweg 4, 2333 CA Leiden, Netherlands \and
    \label{inst:28} Centre Vie dans l’Univers, Faculté des sciences, Université de Genève, Quai Ernest-Ansermet 30, 1211 Genève 4, Switzerland \and
    \label{inst:29} National and Kapodistrian University of Athens, Department of Physics, University Campus, Zografos GR-157 84, Athens, Greece \and
    \label{inst:30} Astrobiology Research Unit, Université de Liège, Allée du 6 Août 19C, B-4000 Liège, Belgium \and
    \label{inst:31} Department of Astrophysics, University of Vienna, Türkenschanzstrasse 17, 1180 Vienna, Austria \and
    \label{inst:32} Institute for Theoretical Physics and Computational Physics, Graz University of Technology, Petersgasse 16, 8010 Graz, Austria \and
    \label{inst:33} Konkoly Observatory, Research Centre for Astronomy and Earth Sciences, 1121 Budapest, Konkoly Thege Miklós út 15-17, Hungary \and
    \label{inst:34} ELTE E\"otv\"os Lor\'and University, Institute of Physics, P\'azm\'any P\'eter s\'et\'any 1/A, 1117 Budapest, Hungary \newpage\and
    \label{inst:35} Lund Observatory, Division of Astrophysics, Department of Physics, Lund University, Box 118, 22100 Lund, Sweden \and
    \label{inst:36} IMCCE, UMR8028 CNRS, Observatoire de Paris, PSL Univ., Sorbonne Univ., 77 av. Denfert-Rochereau, 75014 Paris, France \and
    \label{inst:37} Institut d'astrophysique de Paris, UMR7095 CNRS, Université Pierre \& Marie Curie, 98bis blvd. Arago, 75014 Paris, France \and
    \label{inst:38} Department of Astronomy \& Astrophysics, University of Chicago, Chicago, IL 60637, USA \and
    \label{inst:39} NHFP Sagan Fellow \and
    \label{inst:40} Astrophysics Group, Lennard Jones Building, Keele University, Staffordshire, ST5 5BG, United Kingdom \and
    \label{inst:41} European Space Agency, ESA - European Space Astronomy Centre, Camino Bajo del Castillo s/n, 28692 Villanueva de la Cañada, Madrid, Spain \and
    \label{inst:42} INAF, Osservatorio Astrofisico di Catania, Via S. Sofia 78, 95123 Catania, Italy \and
    \label{inst:43} Institute of Optical Sensor Systems, German Aerospace Center (DLR), Rutherfordstrasse 2, 12489 Berlin, Germany \and
    \label{inst:44} Weltraumforschung und Planetologie, Physikalisches Institut, University of Bern, Gesellschaftsstrasse 6, 3012 Bern, Switzerland \and
    \label{inst:45} Dipartimento di Fisica e Astronomia "Galileo Galilei", Università degli Studi di Padova, Vicolo dell'Osservatorio 3, 35122 Padova, Italy \and
    \label{inst:46} Cavendish Laboratory, JJ Thomson Avenue, Cambridge CB3 0HE, UK \and
    \label{inst:47} Institut fuer Geologische Wissenschaften, Freie Universitaet Berlin, Maltheserstrasse 74-100,12249 Berlin, Germany \and
    \label{inst:48} Institut de Ciencies de l'Espai (ICE, CSIC), Campus UAB, Can Magrans s/n, 08193 Bellaterra, Spain \and
    \label{inst:49} Institut d'Estudis Espacials de Catalunya (IEEC), 08860 Castelldefels (Barcelona), Spain \and
    \label{inst:50} European Space Agency (ESA), European Space Operations Centre (ESOC), Robert-Bosch-Str. 5, D-64293 Darmstadt, Germany \and
    \label{inst:51} HUN-REN-ELTE Exoplanet Research Group, Szent Imre h. u. 112., Szombathely, H-9700, Hungary \and
    \label{inst:52} Institute of Astronomy, University of Cambridge, Madingley Road, Cambridge, CB3 0HA, United Kingdom
}

\date{Received 6 December 2024; accepted XXX}
 
\abstract{Studying the composition of exoplanets is one of the most promising approaches to observationally constrain planet formation and evolution processes. However, this endeavour is complicated for small exoplanets by the fact that a wide range of compositions is compatible with their observed bulk properties. To overcome this issue, we identify triangular regions in the mass-radius space where part of this intrinsic degeneracy is lifted for close-in planets, since low-mass H/He envelopes would not be stable due to high-energy stellar irradiation. Planets in these Hot Water World triangles need to contain at least some heavier volatiles and are therefore interesting targets for atmospheric follow-up observations. We perform a demographic study to show that only few well-characterised planets in these regions are currently known and introduce our CHEOPS GTO programme aimed at identifying more of these potential hot water worlds. Here, we present CHEOPS observations for the first two targets of our programme, TOI-238~b and TOI-1685~b. Combined with TESS photometry and published RVs, we use the precise radii and masses of both planets to study their location relative to the corresponding Hot Water World triangles, perform an interior structure analysis and study the possible lifetimes of H/He and water-dominated atmospheres under these conditions. We find that TOI-238~b lies, at the 1$\sigma$ level, inside the corresponding triangle. While a pure H/He atmosphere would have evaporated after 0.4-1.3~Myr, it is likely that a water-dominated atmosphere would have survived until the current age of the system, which makes TOI-238~b a promising candidate for a hot water world. Conversely, TOI-1685~b lies below the mass-radius model for a pure silicate planet, meaning that even though a water-dominated atmosphere would be compatible both with our internal structure and evaporation analysis, we cannot rule out the planet to be a bare core.}


\maketitle
%
\section{Introduction}
\label{sec:intro}
Understanding how planetary systems form and evolve is one of the key questions in the study of exoplanets. Unfortunately, observational evidence for these processes is quite limited as we can observe only the present-day exoplanet population, thereby capturing a snapshot in time that is further clouded by observational biases. One of the most promising approaches to link this observational picture to planet formation and evolution processes is to study the composition of the observed planets. The presence of water in close-in exoplanets, for instance, can be a sign for these planets having migrated inwards, as we expect water-rich solids to be accreted in the outer part of the protoplanetary disk beyond the ice line. As type I migration, the type of disk migration affecting small planets, is proportional to the planetary mass \citep{Tanaka+2002} and depends on the protoplanetary disk structure, it is expected that close-in planets showcase a variety of water contents, with smaller planets being more water poor than larger ones. This effect is indeed found by recent planetary system formation models \citep{Emsenhuber+2021a,Emsenhuber+2021b}, where a gradient of water content as a function of mass and period is seen at periods lower than $\sim$10 days. In these models, the amount of water in such planets is finally heavily correlated with the respective positions of their formation location, and the protoplanetary disk's ice line location.

In practice, determining the exact composition and more specifically the water mass fraction of an observed exoplanet is a highly degenerate problem, as there are always multiple compositions that can explain the observed mass and radius values of a planet \citep[][]{Valencia+2007,Seager+2007,Rogers+Seager2010}. While many planetary interior models have been developed \citep[e.g.][]{Brugger+2017,Dorn+2017,Acuna+2021,Dorn+Lichtenberg2021,Vazan+2022,Unterborn+2023,Haldemann+2024}, along with statistical methods to infer the interior structure of an observed exoplanet based on classical Bayesian inference algorithms \citep{Dorn+2015,Dorn+2017,Acuna+2021,Haldemann+2024} or machine learning approaches \citep{Baumeister+2020,Haldemann+2023,Baumeister+Tosi2023,Egger+2024}, it is generally not possible to break this intrinsic degeneracy for an individual planet based on density measurements alone, even if the corresponding mass and radius are known very precisely. This makes a direct comparison between the theoretically predicted and the observed exoplanet populations challenging. What could help resolve this problem is the observation of transiting planets at different evolutionary ages, which can statistically constrain the planets' volatile content \citep{Alibert2016}. This approach is especially interesting in light of the upcoming PLATO mission \citep{Rauer+2024}, which will also measure precise stellar ages. Furthermore, additional observational data can help to resolve the degeneracy around the internal structure of exoplanets to some extent. A prominent example of this are upper atmospheric measurements, which have recently been obtained e.g. using the James Webb Space Telescope \citep[JWST;][]{Gardner+2006} for a few super-Earths and sub-Neptunes (e.g. TOI-270~d, \citealp{Benneke+2024,Holmberg+Madhusudhan2024}; K2-18~b, \citealp{Madhusudhan+2023b}; GJ 1214~b, \citealp{Nixon+2024}; GJ~9827~d, \citealp{Piaulet+2024}). While these measurements only constrain the composition of the upper atmosphere at pressures between 1~mbar and 1~bar and the link to lower atmospheric layers and the planetary core remains unclear, jointly modelling the atmosphere and the planetary interior in a self-consistent way can help to reduce the degeneracy to some extent \citep{Guzman+2022,Nixon+2024}.

When we look at planets close to their host star (with orbital periods smaller than $\sim$10 days), part of this intrinsic degeneracy is however lifted without any additional observations. Since they receive a high level of high-energy irradiation, we expect at least part of their atmospheres to be evaporated, an effect that is especially important in the case of H/He atmospheres because of their low mean molecular weight. Indeed, envelopes with mass fractions of less than $\sim$1\% are lost quickly for close-in planets with pure H/He envelopes, within a few tens of Myr for planets with orbital periods $\lesssim$100 days and within a few Myr for planets lighter than 3-5\,$M_{\oplus}$ close to the Neptunian desert \citep{Owen+Wu2017,Lopez2017,Jin+Mordasini2018,kubyshkina_2021mesa}. This means that such planets should either be bare cores or harbour envelopes that make up at least $\sim$1\% of their total mass. Also observationally, we find that most ultra-short period planets have densities compatible with bare rocks \citep{Sanchis-Ojeda+2014,Dai+2021}. This result leads to triangular regions in the mass-radius diagram where it is not possible for close-in planets to exist unless they contain volatiles heavier than H/He. Water being the most abundant of all such volatiles, these planets are promising candidates for so-called water worlds, which in this context we define as planets with water mass fractions of at least a few percent. In the following, we therefore refer to these regions in the mass-radius space as `Hot Water World triangles'.

The existence of such close-in, water-rich planets is of particular interest as this could confirm one of the fundamental predictions of planet formation theory, namely large-scale migration \citep[e.g.][]{Ward1997}. More recently, such water-rich planets have also been proposed as a possible explanation for the radius valley, one of the most heavily debated problems in the exoplanet field to date. This underdensity in the radius distribution of small exoplanets between $\sim$1.5 and $\sim$2 R$_\oplus$, separating super-Earths from sub-Neptunes, was first observed in the California Kepler Survey \citep{Fulton+2017} after already being theoretically predicted independently by multiple groups \citep{Owen+Wu2013,Jin+2014,Lopez+Fortney2014}. The water-rich sub-Neptune hypothesis was first proposed based on mass-radius curves \citep[e.g.][]{Zeng+2019} and later shown to naturally reproduce the observed location of the radius valley from combined formation and evolution simulations \citep{Venturini+2020a,Venturini+2020b,Izidoro+2022,Burn+2024,Venturini+2024}. We note that the location of the valley is also well reproduced by models assuming sub-Neptunes with H/He envelopes \citep[e.g.][]{Owen+Wu2017,gupta_schlichting2019MNRAS.487...24G,Mordasini2020}, which have been matched to radius valley observations in terms of the valley's slope \citep[e.g.][]{VanEylen+2018} and stellar mass dependence \citep{Fulton+Petigura2018,Cloutier+Menou2020,Petigura+2022,Ho+VanEylen2023,Bonfanti+2024,Ho+2024}.

On the observational side, the search for water worlds is a topic of significant interest in the scientific literature at the moment. On the one hand, \cite{Luque+Palle2022} looked at a sample of planets around M~dwarfs and pointed out that many of them lie on the 50\% water line in the mass-radius diagram, suggesting a distinct population of water worlds. More recently and using an updated, larger sample of M~dwarf planets, \cite{Parviainen+2024} argue that a larger sample is needed to test the hypothesis statistically, while \cite{Parc+2024} come to the conclusion that water worlds do not in fact seem to form a distinct population around M~dwarfs. Meanwhile, \cite{Rogers+2023} points out that the properties of these planets can in fact also be explained by H/He dominated atmospheres. On the other hand, the advances in atmospheric observations of small exoplanets in the age of JWST have made it possible to directly search for the presence of water in the atmospheres of these planets. The analysis of recently obtained transmission spectra revealed high atmospheric metallicities for the smaller sub-Neptunes TOI-270~d and GJ~9827~d, while the larger sub-Neptune K2-18~b shows a lower atmospheric metallicity. Although both \cite{Benneke+2024} and \cite{Holmberg+Madhusudhan2024} at least tentatively detect water in the upper atmosphere of TOI-270~d (at 2.5$\sigma$ and 1.6-4.4$\sigma$ respectively), the two works strongly disagree in their interpretation of the planet's nature.\cite{Piaulet+2024} also find water in the atmosphere of GJ~9827~d and conclude that the planet likely is a highly metal-enriched steam world. Meanwhile, \cite{Madhusudhan+2023b} report a non-detection of water in the atmosphere of K2-18~b. Also for other planets, water has been detected using either Hubble, Spitzer or JWST data (e.g. K2-138~c and d, \citealp{Piaulet+2023}). \cite{Moran+2023} note that their detection of water vapour in the atmosphere of the super-Earth GJ~486~b with JWST could also be due to stellar contamination, as GJ~486 is an M~dwarf and therefore has star spots cool enough to contain water. More recent JWST observations by \cite{Mansfield+2024} of the same planet are inconsistent with a water-rich atmosphere, thereby supporting this hypothesis. Moreover, various groups also show that other atmospheric species could act as tracers for liquid water on a planet's surface, such as CO$_2$ \citep{Triaud+2023} or the combination of CH$_4$, CO$_2$ and a non-detection of NH$_3$ \citep{Madhusudhan+2023a}. However, for such models it is important to ensure that a separate, condensed water layer would be stable at the specific atmospheric temperatures and pressures \citep[e.g.][]{Turbet+2020,Pierrehumbert2023,Innes+2023}.

Overall, observational evidence for water worlds remains sparse. With limited observational time on JWST, planets that have to contain at least some heavier volatiles based on their equilibrium temperature, mass and radius are particularly interesting targets. Unfortunately, only a small fraction of the currently known exoplanets fall into the Hot Water World triangles, and for many of the ones that do, multiple contradicting mass or radius values can be found in the literature. For this reason, we initiated a CHEOPS Guaranteed Time Observation (GTO) programme (Hot Water Worlds, CH\_PR140068) that aims to identify new planets located in these sparsely populated regions by using the CHEOPS satellite \citep{Benz+2021,Fortier+2024}.

The exact boundaries of the Hot Water World triangles are defined in more detail in Section \ref{sec:HWW}, where we also perform a demographic study of the known planets that fall into these regions. In Section \ref{sec:targets}, we present observational data for the first two targets of our CHEOPS GTO programme, TOI-238~b and TOI-1685~b, and run both an internal structure analysis and evaporation calculations for the two planets. We finally discuss the limitations of our model in Section \ref{sec:discussion} before providing a summary and drawing conclusions in Section \ref{sec:conclusions}.

\section{Hot Water World triangles}
\label{sec:HWW}
\subsection{Definition}
\label{sec:definition}

\begin{figure*}
    \centering
    \includegraphics[width=0.85\textwidth]{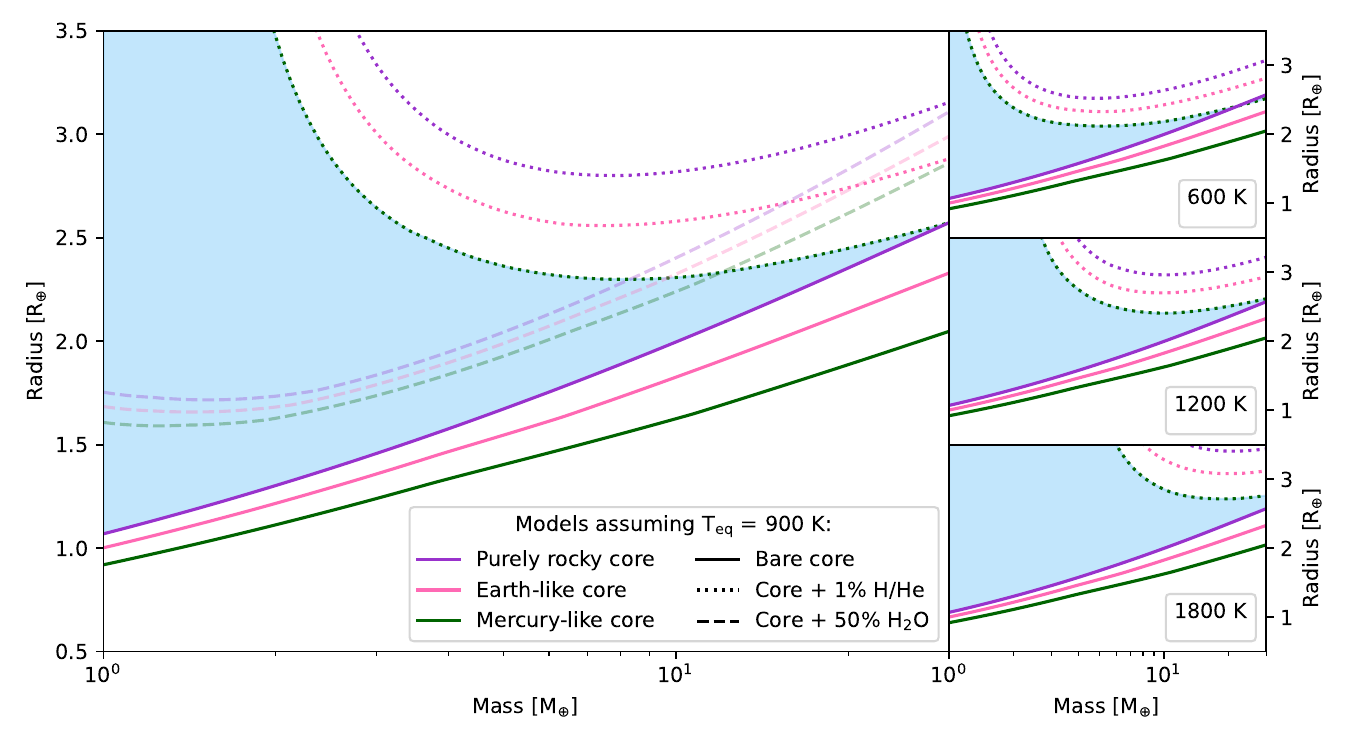}
    \caption{Definition of the Hot Water World triangles (light blue) in the mass-radius space. The left panel shows mass-radius models generated using the BICEPS forward model \citep{Haldemann+2024}, for a fixed equilibrium temperature of 900~K. We generated models for three different core compositions, an iron-less, purely rocky core (in purple), an Earth-like core (33\% inner iron core, 67\% silicate mantle, in pink), and a Mercury-like core (70\% inner iron core, 30\% silicate mantle, in green). For each of these core compositions, we then show three different mass-radius relations, one assuming a bare core (solid lines), one with a 1\% H/He envelope (dotted lines), and one with a 50\% steam atmosphere (dashed lines). The three panels on the right show the temperature dependence of the Hot Water World triangles, with the same mass-radius relations as before now generated for different equilibrium temperatures of 600~K (top), 1200~K (middle), and 1800~K (bottom). See text for further details.}
    \label{fig:triangle_definition}
\end{figure*}

While determining the internal structure of an observed exoplanet based only on its mean density and equilibrium temperature is a highly degenerate problem, part of this degeneracy is lifted for planets close to their host star. According to predictions of atmospheric evaporation theories, we expect at least part of their atmospheres to be evaporated due to the high level of high-energy irradiation they receive \citep[e.g.][]{Lammer2003ApJ...598L.121L,Yelle2004Icar..170..167Y,GMunoz2023A&A...672A..77G,GMunoz2024Icar..41516080G}, an effect that is especially important in the case of H/He atmospheres because of their low mean molecular weight. While the exact amount of H/He lost depends on the specific stellar and planetary properties, \cite{Owen+Wu2017} demonstrate using a minimal analytical model that H/He envelopes with mass fractions of less than $\sim$1\% are lost quickly, with other planetary evolution models showing similar results \citep{Lopez2017,Jin+Mordasini2018}. This leads to regions in the mass-radius space where it is not possible for close-in planets to exist unless their atmospheres contain heavier volatiles, such as water. 

This becomes clear if we look at a mass-radius diagram, shown in Figure~\ref{fig:triangle_definition}. We use the planetary structure model of BICEPS \citep{Haldemann+2024} to generate models for bare cores without a H/He envelope (solid lines) and planets harbouring a 1\% H/He envelope (dotted lines), each for different core compositions. We then consider the most extreme density cases for both scenarios: the purple solid line for the lowest density bare core and the green dotted line for the highest density case of a planet harbouring a 1\% H/He envelope. The triangular region between these two curves is what we define as the Hot Water World triangle corresponding to the chosen equilibrium temperature, highlighted in light blue in the figure. Planets with pure H/He envelopes should not exist in this region, as their envelopes would have quickly been stripped away, but they can be explained by envelope compositions with a higher mean molecular weight (e.g. pure water envelopes shown as dashed lines in Figure~\ref{fig:triangle_definition}). This means that for planets that do lie in these Hot Water World triangles (one for each set of models at a given equilibrium temperature), we can conclude, simply based on their measured masses and radii, that these planets do need to contain heavier volatiles in their envelopes. This makes them especially interesting targets for atmospheric follow-up observations.

Since mass-radius models have a strong temperature dependency, the exact boundaries of this triangle depend on the planet's equilibrium temperature. We note that we limit our analysis to planets with T$_\mathrm{eq}\geq600$~K, which means that potential steam envelopes would be too hot for the water to condense (as e.g. studied by \citealp{Venturini+2024} for the BICEPS forward model used here). For lower equilibrium temperatures, the triangular regions become smaller and eventually even disappear. All mass-radius models with envelopes shown here were generated assuming an intrinsic luminosity of $1 \times 10^{21}$~erg~s$^{-1}$, which is about $2.5$ L$_\oplus$ \citep[with L$_\oplus = 4 \times 10^{20}$~erg~s$^{-1}$; ][]{Kamland2011}. We discuss the limitations of these simplifications (1\% H/He envelopes with a fixed intrinsic luminosity value) in more detail in Section~\ref{sec:discussion_limitations}. We already note, however, that we apply these assumptions just for a first selection of interesting targets, while we later run a more sophisticated evaporation analysis specifically for each observed target.

\subsection{Confirmed planets inside the triangles}
\label{sec:confirmed_planets_in_triangles}

\begin{figure*}
    \centering
    \includegraphics[width=\textwidth,trim={0 0 1.5cm 0},clip]{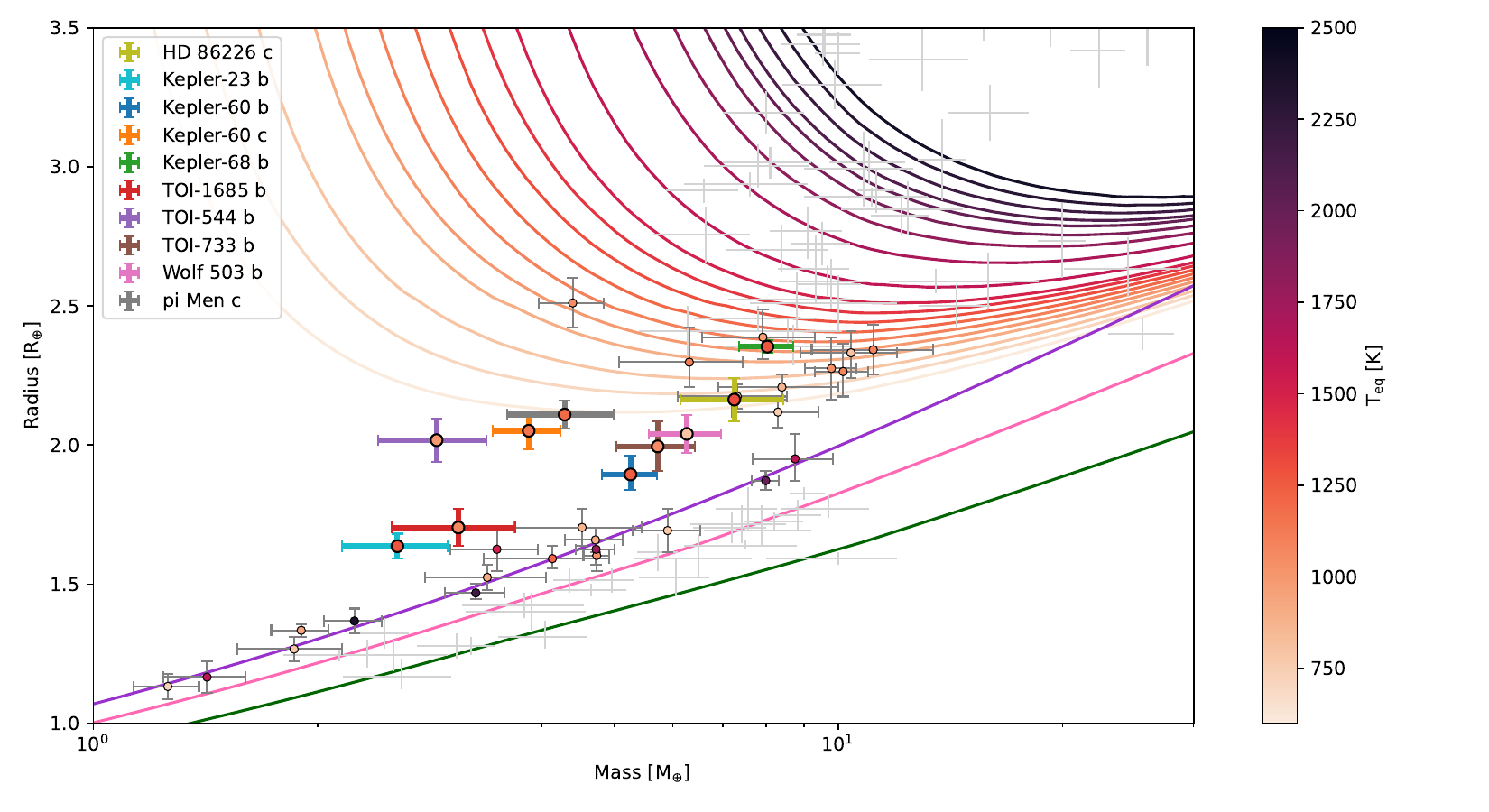}
    \caption{Currently known exoplanets falling inside the Hot Water World triangles. Shown are confirmed planets listed in the PlanetS exoplanet catalogue \citep{Parc+2024, Otegi+2020a}, with a precision in mass of at least 20\% and at least 5\% in radius, with equilibrium temperatures of more than 600~K. The depicted mass-radius curves for bare cores are the same as in Figure~\ref{fig:triangle_definition}, while the ones for planets with a Mercury-like core and a 1\% H/He envelope are colour-coded according to the assumed equilibrium temperature, reaching from 600 up to 2500~K. Planets that fully or partially lie within the Hot Water World triangle corresponding to their equilibrium temperatures are colour-coded according to their equilibrium temperatures, while planets that within two sigma lie fully inside the triangle are additionally marked with bold, colourful error bars. These are HD~86226~c \citep{Teske+2020}, Kepler-23~b \citep{Leleu+2023}, Kepler-60~b and c \citep{Leleu+2023}, Kepler-68~b \citep{Bonomo+2023}, TOI-1685~b \citep{Luque+Palle2022}, TOI-544~b \citep{Osborne+2024}, TOI-733~b \citep{Georgieva+2023}, Wolf~503~b \citep{Polanski+2021}, and $\pi$~Men~c \citep{Damasso+2020}. Planets outside the triangles are marked as grey crosses.}
    \label{fig:demographics}
\end{figure*}

After defining the Hot Water World triangles, we queried the PlanetS exoplanet catalogue \citep[][query from 28 October 2024]{Parc+2024, Otegi+2020a} for already known planets located in these triangular regions. In a first step, we searched for planets with an equilibrium temperature of at least 600~K, a mass of less than 30~Earth masses, and uncertainties of less than 20\% in mass and 5\% in radius, where at least part of their 1$\sigma$ error interval in the mass-radius space lies inside the Hot Water World Triangle corresponding to the planet's equilibrium temperature. These 36~planets are marked with dark grey thin error bars in Figure~\ref{fig:demographics} and will be referred to as Set~1 hereafter. As a next step, we looked for the most promising Hot Water World candidates within this set, which we defined as planets which, at the 2$\sigma$ level, lie fully within the corresponding triangles. The ten planets that this second query yielded (Set~2) are: 
\begin{itemize}
    \item HD~86226~c \citep{Teske+2020}, 
    \item Kepler-23~b \citep{Leleu+2023}, 
    \item Kepler-60~b and c \citep{Leleu+2023}, 
    \item Kepler-68~b \citep{Bonomo+2023}, 
    \item TOI-1685~b \citep{Luque+Palle2022}, 
    \item TOI-544~b \citep{Osborne+2024}, 
    \item TOI-733~b \citep{Georgieva+2023}, 
    \item Wolf~503~b \citep{Polanski+2021}, and
    \item $\pi$~Men~c \citep{Damasso+2020}.
\end{itemize} 
They orbit a wide variety of different host stars, with masses ranging from 0.5 to 1.1~M$_\odot$ and ages between 1.3 and 11.0~Gyr. In Figure~\ref{fig:demographics}, they are labelled and marked with thicker, coloured error bars. The markers for both sets of planets are colour-coded according to their equilibrium temperatures. The upper boundaries of the corresponding Hot Water World triangles are also colour-coded according to the assumed planetary equilibrium temperature using the same colour map. All other confirmed planets that fulfil the same criteria but lie, within 1$\sigma$, fully outside their corresponding Hot Water World triangles are marked as light grey crosses. This group contains 122~planets (Set~3).

Many of the planets contained in Set~1, but not Set~2, are so-called low density super-Earths as identified e.g. by \cite{Castro-Gonzalez+2023}. These planets lie between the theoretical models for bare cores with a purely rocky and an Earth-like composition in the mass-radius space. Their mean density can therefore be explained by a volatile-rich atmosphere, but also by a bare core that is iron-poor.

Similar arguments as we make here when defining the Hot Water World triangles have been made in the literature for one of the planets in Set~2, $\pi$~Men~c, by \cite{GarciaMunoz+2020} and \cite{GarciaMunoz+2021}. They additionally strengthen their argument of a high mean molecular weight atmosphere through HST observations of $\pi$~Men~c, reporting a 3.4$\sigma$ detection of escaping C~II ions, as well as a non-detection of H~I atoms. Another planet, GJ~9827~d, which is not part of Set~2 because of its radius uncertainty of >5\%, has recently been observed with JWST. \cite{Piaulet+2024} also argue that a pure H/He atmosphere would likely have evaporated and report the detection of a water-dominated atmosphere.

\section{The two first targets of the Hot Water Worlds CHEOPS programme}
\label{sec:targets}

\begin{figure*}
    \centering
    \includegraphics[width=\textwidth]{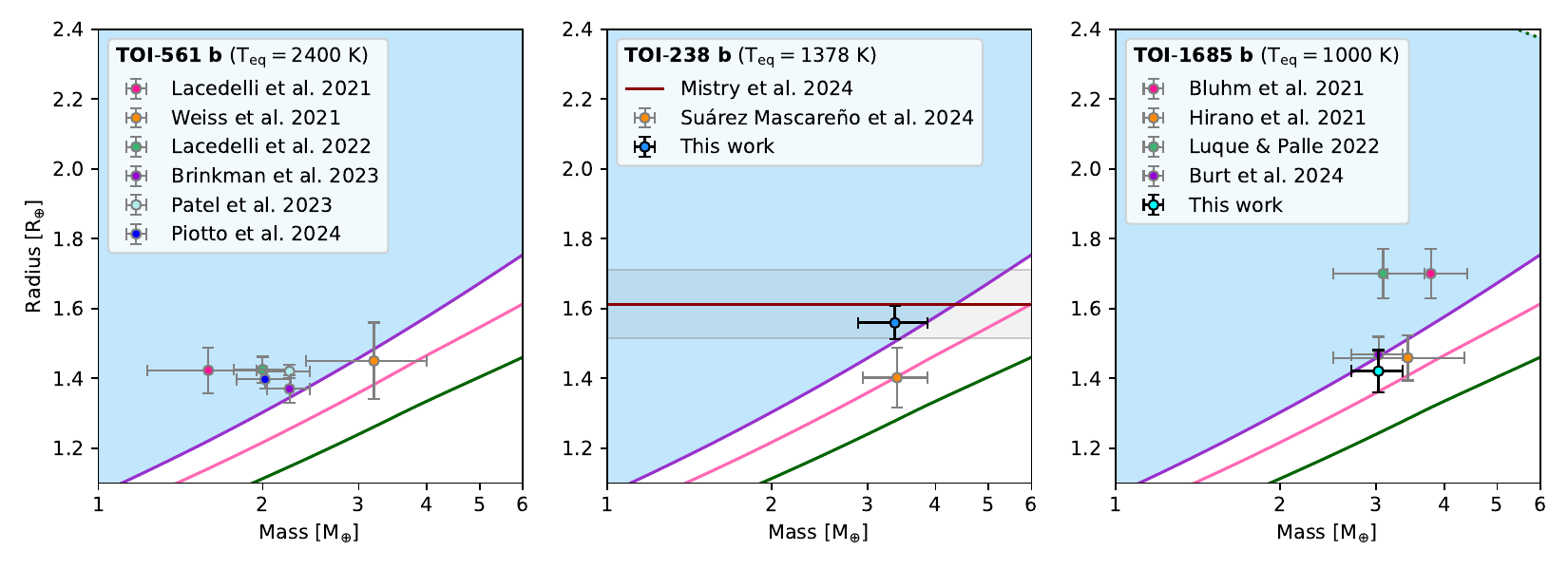}
    \caption{Mass-radius diagrams for TOI-561~b (left), TOI-238~b (middle) and TOI-1685~b (right), showcasing the corresponding Hot Water World triangles. The mass-radius models are the same as in Figure~\ref{fig:triangle_definition} but for each planet's specific equilibrium temperature. The depicted literature values are summarised in Table~\ref{tab:literature_values}.}
    \label{fig:triangle_obs}
\end{figure*}

In Section~\ref{sec:HWW}, we have defined the boundaries of our Hot Water World triangles and shown that only few currently known and well-characterised planets fall inside these triangles. Furthermore, follow-up observations for planets that lie fully or partially inside the triangles can lead to drastically different mass and radius values, which changes the planets' locations in the mass-radius space and therefore whether they are located inside or outside the corresponding triangles. One example of this is TOI-561~b, which has already been studied extensively with CHEOPS \citep{Lacedelli+2022,Patel+2023,Piotto+2024}. The top panel of Figure~\ref{fig:triangle_obs} shows this planet in the mass radius space, with the respective Hot Water World Triangle marked in light blue. It becomes obvious that the planetary parameters in the literature vary over quite a wide range, in the case of TOI-561~b mostly the mass. The situation is similar for other planets in sets~1 and 2 defined in the previous section, with conflicting literature values either in mass or radius. This highlights the importance of follow-up observations especially for small planets, not only to improve the precision but also to confirm the accuracy of the determined mass and radius values.

We are therefore following up on the radii of confirmed planets and planetary candidates that could potentially lie inside the triangles as part of a CHEOPS GTO programme (Hot Water Worlds, CH\_PR140068), with the aim of increasing the sample size of this very interesting set of planets. In the following, we present CHEOPS observations for the first two targets of this programme, TOI-238~b and TOI-1685~b. We perform a stellar analysis for both targets and analyse the new CHEOPS data jointly with the available TESS data. After deriving updated radii and masses (based on the RV semi-amplitudes reported in \citeauthor{Mascareno+2024} \citeyear{Mascareno+2024} and \citeauthor{Burt+2024} \citeyear{Burt+2024} as well as our newly derived stellar masses), we study the internal structure of both planets and run a detailed evaporation analysis.

\begin{table*}
\renewcommand{\arraystretch}{1.4}
\caption{Reported literature values for the planetary masses and radii of TOI-561~b, TOI-238~b, and TOI-1685~b, as well as the corresponding stellar masses and radii.}
\centering
\begin{tabular}{lccccc}
\hline\hline
Planet & Planetary radius [R$_\oplus$] & Planetary mass [M$_\oplus$] & Stellar radius [R$_\odot$] & Stellar mass [M$_\odot$] & Source\\
\hline
\multirow{6}{*}{TOI-561~b} & 1.423 $\pm$ 0.066 & 1.59 $\pm$ 0.36 & 0.849 $\pm$ 0.007 & 0.785 $\pm$ 0.018 & [1] \\
 & 1.45 $\pm$ 0.11 & 3.2 $\pm$ 0.8 & 0.832 $\pm$ 0.019 & 0.805 $\pm$ 0.030 & [2] \\
 & 1.425 $\pm$ 0.037 & 2.00 $\pm$ 0.23 & 0.843 $\pm$ 0.005 & 0.806 $\pm$ 0.036 & [3] \\
 & 1.37 $\pm$ 0.04 & 2.24 $\pm$ 0.20 & 0.832 $\pm$ 0.019 & 0.805 $\pm$ 0.030 & [4] \\
 & 1.42 $\pm$ 0.02 & 2.24 $\pm$ 0.20 & 0.843 $\pm$ 0.005 & 0.806 $\pm$ 0.036 & [5] \\
 & 1.397 $\pm$ 0.027 & 2.02 $\pm$ 0.23 & 0.843 $\pm$ 0.005 & 0.806 $\pm$ 0.036 & [6] \\
\hline
\multirow{3}{*}{TOI-238~b} & $1.612^{+0.096}_{-0.099}$ &  --  & $0.750^{+0.025}_{-0.026}$ & $0.788^{+0.047}_{-0.048}$ & [7] \\
 & 1.402 $\pm$ 0.086 & 3.40 $\pm$ 0.46 & 0.733 $\pm$ 0.015 & 0.790 $\pm$ 0.022 & [8] \\
 & 1.559 $\pm$ 0.047 & 3.37 $\pm$ 0.49 & 0.756 $\pm$ 0.007 & $0.781^{+0.040}_{-0.039}$ & [13] \\
\hline
\multirow{6}{*}{TOI-1685~b} & 1.70 $\pm$ 0.07 & 3.78 $\pm$ 0.63 & 0.492 $\pm$ 0.015 & 0.495 $\pm$ 0.019 & [9] \\
 & 1.459 $\pm$ 0.065 & 3.43 $\pm$ 0.93 & 0.459 $\pm$ 0.013  & 0.460 $\pm$ 0.011 & [10] \\
 & 1.70 $\pm$ 0.07 & 3.09 $\pm$ 0.59 & 0.492 $\pm$ 0.015 & 0.495 $\pm$ 0.019 & [11] \\
 & $1.468^{+0.050}_{-0.051}$ & $3.03^{+0.33}_{-0.32}$ & 0.4555 $\pm$ 0.0128 & 0.454 $\pm$ 0.018 & [12] \\
 & 1.421 $\pm$ 0.060 & $3.07^{+0.34}_{-0.33}$ & 0.462 $\pm$ 0.013 & 0.466 $\pm$ 0.019 & [13] \\
\hline
\end{tabular}
\label{tab:literature_values}
\tablebib{
[1]~\cite{Lacedelli+2021}; [2]~\citet{Weiss+2021}; [3]~\citet{Lacedelli+2022}; [4]~\cite{Brinkman+2023}; [5]~\citet{Patel+2023}; [6]~\citet{Piotto+2024}; [7]~\cite{Mistry+2024}; [8]~\citet{Mascareno+2024}; [9]~\citet{Bluhm+2021}; [10]~\cite{Hirano+2021}; [11]~\citet{Luque+Palle2022}; [12]~\citet{Burt+2024}; [13]~This work.
}
\end{table*}
\renewcommand{\arraystretch}{1}


\subsection{TOI-238 b}
\label{sec:TOI-238b}
TOI-238~b orbits a bright K~dwarf on a 1.27~day orbit. The planet was first identified as a transiting planet candidate by TESS and validated by \cite{Mistry+2024}, who reported a radius of $1.61^{+0.09}_{-0.10}$~R$_\oplus$. \cite{Mascareno+2024} obtained ESPRESSO and HARPS RVs that they jointly fit with the available TESS data, reporting a mass of $3.40^{+0.46}_{-0.45}$~M$_\oplus$ and a radius of $1.402^{+0.084}_{-0.086}$~R$_\oplus$, more than 2$\sigma$ below the value reported by \cite{Mistry+2024}. While the first radius value would make TOI-238~b a very promising potential Hot Water World, with this second set of parameters the planet would lie below the mass-radius model for a purely rocky core and thereby outside of the corresponding triangle, as shown in the middle panel of Figure~\ref{fig:triangle_obs}. \cite{Mascareno+2024} also discovered a second planet in the system, the sub-Neptune TOI-238~c with an orbital period of 8.47~days, which they identified in both the radial velocity data and TESS transit photometry.

\subsubsection{Host star characterisation}\label{sec:starTOI238} 
\begin{table*}
\renewcommand{\arraystretch}{1.4}
\caption{Stellar properties of TOI-238 and TOI-1685.}
\centering
\begin{tabular}{lcrcr}
\hline\hline
\multicolumn{1}{l}{} & \multicolumn{2}{c}{TOI-238} & \multicolumn{2}{c}{TOI-1685} \\
Parameter & Value & Ref. & Value & Ref. \\
\hline
$\alpha$ (J2000) & 23h 16m 55.52s & [1] & 04h 34m 22.50s & [1]\\
$\delta$ (J2000) & -18$^\circ$ 36' 23.92'' & [1] & +43$^\circ$ 02' 14.69'' & [1]\\
G mag & 10.4901 $\pm$ 0.0028 & [1] & 12.2846 $\pm$ 0.0028 & [1]\\
Spectral type & K2V & [2] & M3V & [3]\\
T$_\textrm{eff}$ [K] & 5059 $\pm$ 89 & [2] & 3470 $\pm$ 102 & [4]\\
log g [cgs] & 4.58 $\pm$ 0.05 & [2] & 4.68 $\pm$ 0.11 & [4]\\
$v\sin{i}$ [km s$^{-1}$] & 2.3 $\pm$ 0.3 & [4] & $<2.0$ & [3]\\
$[\mathrm{Fe/H}]$ [dex] & -0.114 $\pm$ 0.051 & [2] & 0.01 $\pm$ 0.12 & [4]\\
$[\mathrm{Mg/H}]$ [dex] & -0.12 $\pm$ 0.06 & [2] & 0.04 $\pm$ 0.17 & [4]\\
$[\mathrm{Si/H}]$ [dex] & -0.09 $\pm$ 0.06 & [2] & 0.02 $\pm$ 0.17 & [4]\\
R$_\star$ [R$_\odot$] & 0.756 $\pm$ 0.007 & [4] & 0.462 $\pm$ 0.013 & [4]\\
M$_\star$ [M$_\odot$] & $0.781_{-0.039}^{+0.040}$ & [4] & 0.466 $\pm$ 0.019 & [4]\\
t$_\star$ [Gyr] & $7.9_{-5.2}^{+4.5}$ & [4] & 1.4 $\pm$ 0.1 & [4]\\
\hline
\end{tabular}
\label{tab:stellar_params}
\tablebib{
[1]~\cite{GaiaCollab2021}; [2]~\citet{Mascareno+2024}; [3]~\citet{Bluhm+2021}; [4]~This work.
}
\end{table*}
\renewcommand{\arraystretch}{1}

We adopted the spectroscopic parameters that were derived and presented in \citet[][]{Mascareno+2024} using the ARES+MOOG methodology \citep[e.g.][]{Sousa+2021} on a combined ESPRESSO spectrum of TOI-238. The $v\sin{i}$ value was re-derived in this work using the same combined spectrum and compared with a synthetic spectrum using the ’spectroscopy made easy’ (SME) code \citep[][]{Piskunov+2017} for which we get a value of 2.3 $\pm$ 0.3 km/s. For the elemental abundances of Mg and Si of TOI-238, we use the values derived by \cite{Mascareno+2024}.

To determine the radius of TOI-238, we used a Markov-Chain Monte Carlo (MCMC) modified infrared flux method \citep{Blackwell1977,Schanche2020}. Priors from our spectral analysis were used to constrain stellar atmospheric models from three catalogs \citep{Kurucz1993,Castelli2003,Allard2014} from which spectral energy distributions (SEDs) were constructed. We computed synthetic photometry from these SEDs that were compared to observed broadband photometry in the following bandpasses: {\it Gaia} $G$, $G_\mathrm{BP}$, and $G_\mathrm{RP}$, 2MASS $J$, $H$, and $K$, and \textit{WISE} $W1$ and $W2$ \citep{Skrutskie2006,Wright2010,GaiaCollaboration2022} in order to calculate the bolometric flux of TOI-238. Using the Stefan-Boltzmann law, we derived the stellar effective temperature and angular diameter that we converted to the stellar radius using the offset-corrected \textit{Gaia} parallax \citep{Lindegren2021}. We conducted a Bayesian modeling averaging of the \textsc{atlas} \citep{Kurucz1993,Castelli2003} and \textsc{phoenix} \citep{Allard2014} catalogs to remove stellar atmospheric model uncertainties when deriving the stellar radius.

We computed the isochronal mass $M_{\star}$ and age $t_{\star}$ of TOI-238 using two different stellar evolutionary models, by inputting $T_{\mathrm{eff}}$, [Fe/H], and $R_{\star}$ along with their error bars. In detail, we interpolated the input set within pre-computed grids of PARSEC\footnote{\textsl{PA}dova and T\textsl{R}ieste \textsl{S}tellar \textsl{E}volutionary \textsl{C}ode: \url{http://stev.oapd.inaf.it/cgi-bin/cmd}} v1.2S \citep{marigo2017} via the isochrone placement algorithm \citep{bonfanti2015,bonfanti2016}. We further constrained the convergence thanks to the available $v\sin{i}$ estimate by coupling the isochrone fitting with the gyrochronological relation of \citet{barnes2010} as detailed in \citet{bonfanti2016}. This way, we derived a first set of mass and age values. A second set of outcomes was inferred with the Code Liègeois d'Évolution Stellaire \citep[CLES;][]{scuflaire2008} that generates an `on-the-fly' evolutionary track following a Levenberg-Marquadt minimisation scheme \citep{salmon2021}.
\citet{bonfanti2021} outlines the $\chi^2$-based criterion we followed to check the consistency of the two respective pairs of outcomes and describes how we merged the results to finally obtain $M_{\star}=0.781_{-0.039}^{+0.040}\,M_{\odot}$ and $t_{\star}=7.9_{-5.2}^{+4.5}$\,Gyr.

\subsubsection{Observations and data analysis} 
\label{sec:obs_238}
\begin{figure*}
    \centering
    \includegraphics[width=0.8\textwidth]{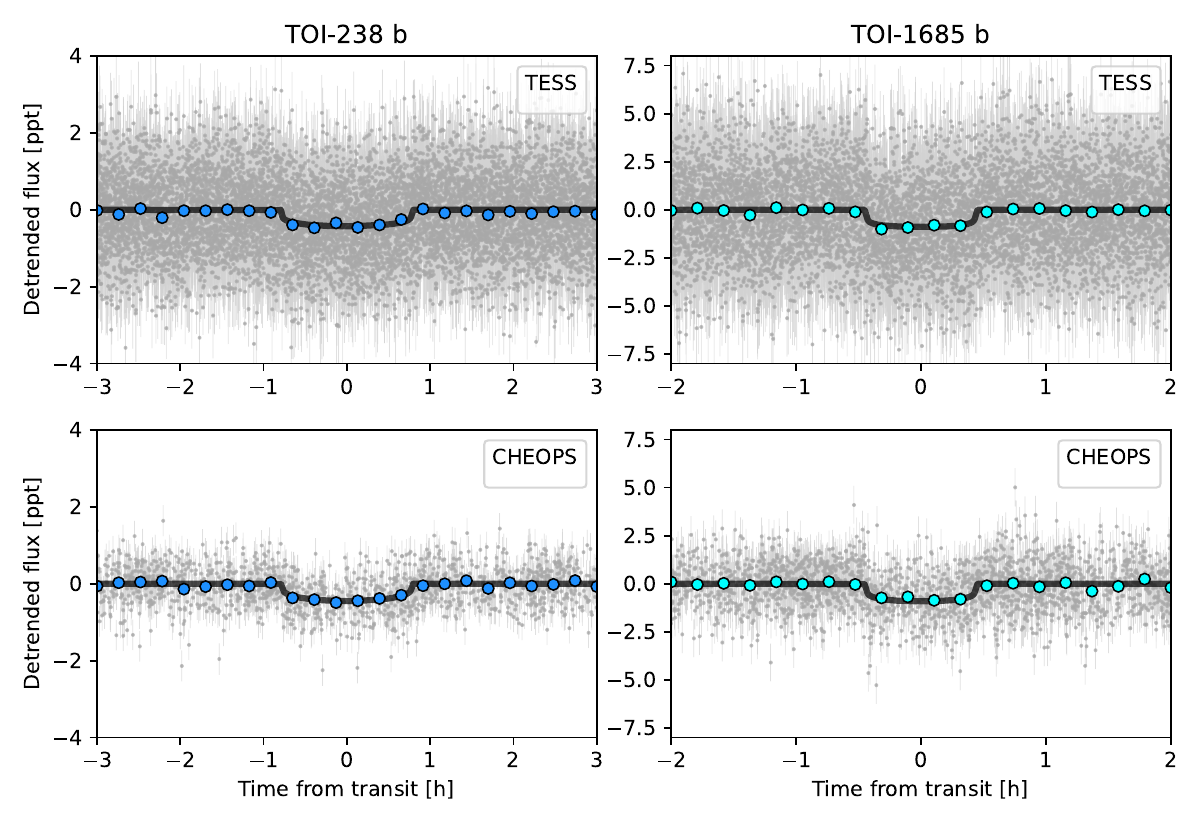}
    \caption{Detrended TESS (top row) and CHEOPS (bottom row) light curves for TOI-238~b (left) and TOI-1685~b (right), phase-folded to the orbital periods of the respective planets. All panels show relative flux normalised to 0.}
    \label{fig:phasefolded_transits}
\end{figure*}

\begin{figure}
    \centering
    \includegraphics[width=\linewidth]{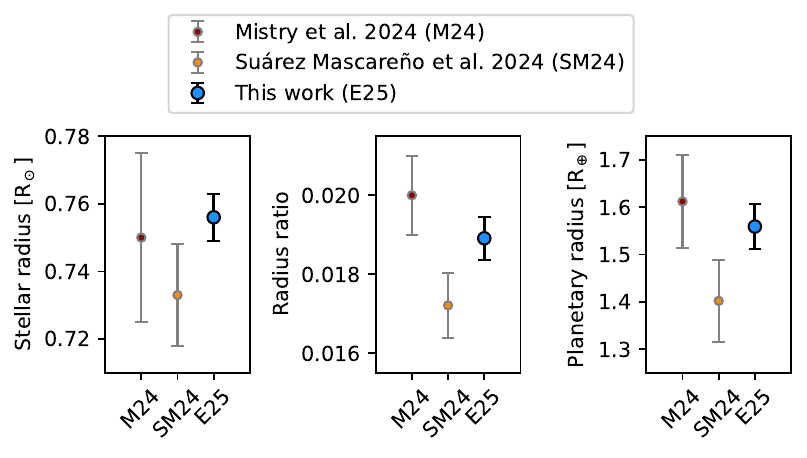}
    \caption{Comparison of the different values in the literature for the stellar radius, radius ratio and planetary radius of TOI-238~b.}
    \label{fig:radius_comp_238}
\end{figure}

\begin{table*}
\renewcommand{\arraystretch}{1.4}
\caption{Posterior distributions of the planetary parameters for TOI-238~b and TOI-1685~b. RV semi-amplitudes from \cite{Mascareno+2024} and \cite{Burt+2024} respectively.}
\centering
\begin{tabular}{lcc}
\hline\hline
Parameter & TOI-238~b & TOI-1685~b\\
\hline
Epoch, $t_0$ [BJD-2457000]  & 2460205.7524 $\pm$ 0.0011 & 2458816.22570 $\pm$ 0.00071 \\
Period, $P$ [d] & 1.2730990 $\pm$ 0.0000016 & 0.66913923 $\pm$ 0.00000039 \\
Radius ratio, $R_p/R_s$ & 0.01891 $\pm$ 0.00055 & 0.02818 $\pm$ 0.00086 \\
Impact parameter, $b$ & 0.23 $\pm$ 0.12 & 0.31 $\pm$ 0.19 \\
Semi-major axis, $a$ [AU] & 0.02120 $\pm$ 0.00038 & 0.01164 $\pm$ 0.00097 \\
Semi-major axis over stellar radius, $a/R_s$ & 6.03 $\pm$ 0.11 & 5.41 $\pm$ 0.46 \\
Transit duration, $t_{14}$ [h] & 1.587 $\pm$ 0.047 & 0.904 $\pm$ 0.019 \\
Radius, $R_p$ [R$_\oplus$] & 1.559 $\pm$ 0.047 & 1.421 $\pm$ 0.060 \\
Incident flux, $S_p$ [kW\,m$^{-2}$] & 1025 $\pm$ 86 & 290 $\pm$ 75 \\
Equilibrium temperature, $T_{eq}$ [K] & 1378 $\pm$ 30 & 1000 $\pm$ 59 \\
Impact parameter, $b$ & 0.23 $\pm$ 0.12 & 0.31 $\pm$ 0.19 \\
\hline
RV semi-amplitude, $K$ [m\,s$^{-1}$] & 2.36 $\pm$ 0.32 & 3.76$^{+0.39}_{-0.38}$\\
Mass $M_p$ [M$_\oplus$] & 3.37 $\pm$ 0.49 & 3.07$^{+0.34}_{-0.33}$\\
\hline
\end{tabular}
\label{tab:planetary_params}
\end{table*}
\renewcommand{\arraystretch}{1}

TOI-238 was observed by TESS in sectors~2 (September 2018), 29 (September 2020) and 69 (September 2023), all at two-minute cadence. We further observed five transits of TOI-238~b with CHEOPS between July and September 2023, for a total of 39.6~hours and as part of the GTO. A CHEOPS observation log can be found in Table~\ref{tab:file_keys}, while the undetrended light curves are shown in Figure~\ref{fig:undetrended_CHEOPS_238}.

To jointly model the available TESS and CHEOPS data, we used the publicly available code \texttt{chexoplanet}\footnote{\url{https://github.com/hposborn/chexoplanet}}, which makes use of the \texttt{exoplanet} library \citep{Foreman-Mackey+2021}. For the available TESS sectors, we used the PDCSAP photometry with additional long-timescale trends removed using a cubic spline with in-transit data masked and breakpoints spaced every 0.9~days. Such trends can either be systematic or caused by stellar rotation. For the CHEOPS photometry, we used PIPE\footnote{\url{https://github.com/alphapsa/PIPE}} \citep{Brandeker+2024} to extract PSF photometry, which helps to reduce contamination by background stars, cosmic ray hits, or hot pixels. 

Additionally, \texttt{chexoplanet} models the impact of systematics by modelling the linear and quadratic decorrelation of the flux with respect to various hyperparameters. This is done by fitting each individual CHEOPS light curve using all available hyperparameter timeseries (x and y centroid position, first three aliases of the cosine and sine of the roll-angle, on-board temperature, background flux, major residuals of the PSF fit, time). Bayesian model comparison then allows us to determine which of these hyperparameters improve the model (Bayes Factor >1). As a second step, we then determine the similarity of these decorrelation parameters across all observations using a leave-one-out comparison. The resulting series of linear and quadratic parameters, which modify the CHEOPS flux according to the variation of the corresponding hyperparameter, is then co-modelled with the transits. Furthermore, \texttt{chexoplanet} co-models shorter-frequency variations of the flux as a function of the spacecraft roll angle. This modulation is caused by the combination of the rotating field of view of CHEOPS due to its nadir-locked orbit around the Earth and the strongly asymmetric PSF of each background star \citep{Benz+2021}. The roll angle pattern is described in more detail in \cite{Lendl+2020} and \cite{bonfanti2021}. These modulations are modelled simultaneously with the exoplanet transits using a cubic spline. We chose a breakpoint spacing of 9~degrees as this is more than double the minimum observing cadence (3.6~degrees) but still enables rapid changes with roll angle to be modelled.

The TESS and CHEOPS transit signals are then co-modelled with the CHEOPS detrending described above. The \texttt{exoplanet} package is used for the transit models, setting circular orbits and normal priors centred on the theoretical limb-darkening coefficients derived by \cite{Claret2017} for TESS and \cite{Claret2021} for CHEOPS. We used a broad log-normal prior for the planetary radius ratio, the prior of \cite{Espinoza2018} for the impact parameter, and broad log-normal priors for jitter terms accounting for additional white noise for each of the two instruments separately. Normal priors were used for timing and orbital periods, with $\sigma$ set using the value from the TOI catalogue.

We finally arrive at a radius ratio of 0.01891$\pm$0.00055, which together with the stellar radius determined in the previous section gives us a planetary radius of $1.559\pm0.047$~R$_\oplus$. This agrees very well with the value found by \cite{Mistry+2024}, but lies $\sim$1.8$\sigma$ (and even $\sim$3.3$\sigma$ using our own smaller uncertainties) above the median value from \cite{Mascareno+2024}. The discrepancies with the radius value derived by \cite{Mascareno+2024} are caused both by differences in the stellar radius as well as the radius ratio (see Figure~\ref{fig:radius_comp_238}). Using the RV semi-amplitude derived by \cite{Mascareno+2024} and the stellar mass derived in the previous section, we find a mass of $3.37\pm0.49$~M$_\oplus$ for TOI-238~b. The resulting posteriors for the planetary parameters are summarised in the left column of Table~\ref{tab:planetary_params}, while the resulting photometry fits are shown in the left panels of Figure~\ref{fig:phasefolded_transits}.

\subsubsection{Internal structure analysis} 
\label{sec:int_struc_238}

\begin{figure*}
    \centering
    \includegraphics[width=\textwidth]{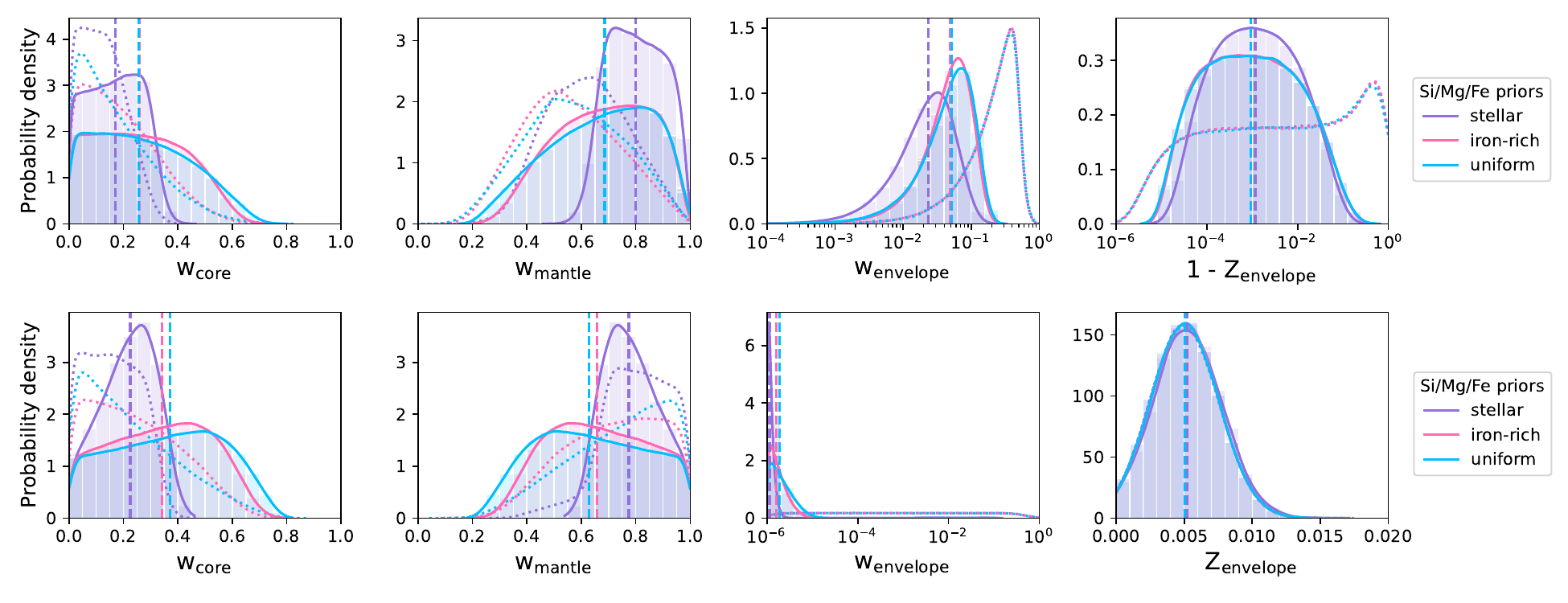}
    \caption{Results of the internal structure analysis for TOI-238~b. Depicted are the posterior distributions for the mass fractions of the inner core, the mantle and the volatile layer with respect to the whole planet, as well as the water mass fraction in the volatile layer, $\mathrm{Z}_\mathrm{envelope}$ (from left to right). The top row (case A) was generated for a prior that would be in agreement with a water-rich composition, e.g. if the planet formed outside the iceline, while the bottom row (case B) assumes that water was not readily available for accretion during planet formation. The three colours depict different assumptions for the planetary Si/Mg/Fe ratios: stellar (purple), iron-enriched (pink), and uniformly sampled without considering the stellar ratios (blue). The dashed vertical lines show the median of each distribution, while the dotted lines show the respective priors.}
    \label{fig:int_struct_238}
\end{figure*}

As a next step, we investigate the internal structure of TOI-238~b. Looking at the planet's location in a mass radius diagram (see middle panel of Figure~\ref{fig:triangle_obs}), we find that TOI-238~b lies above the model for a purely rocky, iron-free bare core at the 1$\sigma$ level, which means that it must contain at least some volatiles other than H/He (since it also lies below the upper boundary of the triangle). Furthermore, from a planet formation point of view it is unlikely that a planet would only accrete silicon and magnesium during its formation but no iron, strengthening this claim even further \citep[e.g.][]{Thiabaud+2015}. While we should consider that TOI-238 has, with [Fe/H]~$=-0.114\pm0.051$, a lower metallicity than the Sun and therefore also might host planets that are less rich in iron compared to the planets in our solar system \citep[e.g.][]{Thiabaud+2015,Adibekyan+2021,Michel+2020}, it is still reasonable to expect TOI-238~b to at least contain a small amount of iron.

For a more detailed analysis of TOI-238~b's internal structure, we run the interior modelling framework \texttt{plaNETic}\footnote{\url{https://github.com/joannegger/plaNETic}} \citep{Egger+2024}, which is based on the planetary structure model of the BICEPS code \citep{Haldemann+2024} and models an observed exoplanet as a layered structure combining an inner iron core with up to 19\% of sulphur, a mantle consisting of oxidised silicon, magnesium and iron, and a volatile layer made up of uniformly mixed water and H/He. While most other interior structure modelling frameworks use Markov chain Monte Carlo algorithms for the inference of the internal structure parameters \citep[e.g.][]{Dorn+2017,Acuna+2021,Haldemann+2024}, \texttt{plaNETic} instead leverages the idea of training neural networks on data generated with the forward model, which are then used in a full-grid accept-reject sampling scheme \citep[see also][where the preliminary version of the code was first introduced]{Leleu+2021}. This procedure significantly speeds up the inference computation: Once trained, the neural network computes the transit radius of a sampled structure $\sim$50~000 times faster than the forward model would, but still gives very accurate results. 

Some other neural network based frameworks such as \cite{Baumeister+Tosi2023} and \cite{Haldemann+2023} replace the full inference scheme with a neural network instead of just the forward model, which leads to an even more significant decrease of the necessary computation time. However, contrary to those frameworks, the \texttt{plaNETic} method allows to freely modify the chosen priors without having to retrain the neural networks and reduces convergence problems that tend to occur if the chosen forward model is too physically complex. Moreover, it also allows for an easy and direct test of how well the neural network performs for a given planet by just re-running a few structures from the sampled posterior distributions with the full forward model.

Here, we use a set of six different priors for the inferred interior structure parameters in order to take into account different formation scenarios and star-planet relations. On the one hand, we consider two different scenarios for the water content of the planet, one compatible with a water-rich composition inspired by a formation location outside of the iceline, and one assuming a water-poor composition in agreement with the planet forming inside the iceline and only accreting water through the accreted gas. In the first case, this means that the mass fractions of the inner core, silicate mantle and water are sampled uniformly from the simplex on which they add up to unity. The sampled water is then uniformly mixed with the accreted H/He to form the volatile layer. Meanwhile, in the second case, the water mass fraction in the envelope is sampled from a Gaussian prior with mean 0.5\% and standard deviation 0.25\%. On the other hand, we use three different priors for the planetary Si/Mg/Fe ratios, one assuming they match those of the host star exactly \citep[e.g.][]{Thiabaud+2015}, one assuming the planet to be enriched in iron \citep{Adibekyan+2021}, and one where the molar fractions of Si, Mg and Fe in the bulk of the planet are sampled uniformly from the simplex where they add up to one, with an upper limit of 0.75 for the molar fraction of iron. These priors are explained in more detail in \cite{Egger+2024}.

The posterior distributions for the most important internal structure parameters of TOI-238~b are visualised in Figure~\ref{fig:int_struct_238}, while a full summary of all parameters is provided in Table~\ref{tab:internal_structure_results_238}. For a water-rich prior (top row of Figure~\ref{fig:int_struct_238}), we infer envelope mass fractions of a few percent relative to the total planet mass and envelope water mass fractions of almost 100\%. If we assume that the planet could only have accreted water through the accreted gas (bottom row), the mass fractions of the almost pure H/He envelopes are of the order of $10^{-6}$, which would be evaporated quickly and would not be stable over longer timescales. These inferred internal structure properties are in agreement with those expected for planets of this type and follow the same trends with respect to the planetary radius, envelope mass fraction and equilibrium temperature as the planets in Set~2 identified in Section~\ref{sec:confirmed_planets_in_triangles}. This was verified by running a simplified internal structure analysis for the planets in Set~2, assuming solar host star properties and Si/Mg/Fe ratios for each planet.

\texttt{plaNETic} also gives a posterior distribution of the planet's intrinsic luminosity, derived for each sampled structure from its core and envelope mass fractions and the stellar age according to the fit from \cite{Mordasini2020}. For TOI-238~b, this leads to an estimate of $\log_{10}\left(\frac{L}{1\textrm{erg s}^{-1}}\right) = 20.4^{+0.5}_{-0.2}$ for the intrinsic luminosity.

\subsubsection{Evolution of hydrogen-helium atmospheres} 
\label{sec:evap_H/He}
To test whether TOI-238\,b could sustain a hydrogen-dominated atmosphere until its age of $7.9_{-5.2}^{+4.5}$\,Gyr (see Table~\ref{tab:stellar_params}), we employ the atmospheric evaporation rates based on the large grid of upper atmosphere models presented in \citet{kubyshkina2018_grid}. The hydrodynamic model used in this study represents a 1D model of pure hydrogen atmospheres accounting for photoionisation heating and basic hydrogen chemistry. The grid consists of 10235 models of hydrogen-dominated atmospheres around planets spanning a wide range of planetary and stellar host parameters, such as planetary mass, radius, orbital separation, stellar mass, and stellar X-ray and extreme ultraviolet (XUV) irradiation levels. Details of the parameters distribution of the model planets can be found in \citet{kubyshkina2021_grid}.
Interpolation between the tabulated values of the atmospheric mass loss rates resolved from the grid models allows us to calculate these rates for any given planetary parameters within the grid range, and, therefore, to track the evolution of planetary atmospheres driven by the evaporation. 
To relate the atmospheric mass and the planetary radius throughout the evolution, we employ MESA models \citep[Modules for Experiments in Stellar Astrophysics; ][]{paxton2018} as described in \citet{kubyshkina_2021mesa}.

Besides the primordial parameters, three factors have a significant impact on the atmospheric evaporation history: the mass of the planet, the orbital separation, and the XUV evolution of the host star. As we are interested in the maximum atmospheric lifetime, we employ the planetary and stellar parameters that minimise the atmospheric evaporation rates within the observational uncertainties for our simulations. Thus, for the mass of TOI-238\,b, we adopt 3.86\,$\textrm{M}_\oplus$, corresponding to the upper limit given in Table~\ref{tab:planetary_params}. The primordial H/He atmospheres of close-in planets in this mass range do not add considerably to the planet's mass. Therefore, we assume that this mass is concentrated within the solid part of the planet, below the rocky core radius. %
The semi-major axis, $a$, is well constrained by the observations; therefore, we employ the median value of 0.02120\,AU and assume that $a$ did not change since the time of protoplanetary disk dispersal.

The most uncertain parameter is the evolution of the stellar XUV luminosity, which can spread largely for different stars at ages below $\sim$1\,Gyr. This depends on the initial stellar rotation period: slower rotating stars emit less in the XUV band. However, at later ages, the XUV evolution pathways of stars born with different rotation rates converge, and the rotation history cannot generally be defined from the present-day stellar parameters. For the TOI-238 system, the age estimation is rather broad, which does not allow us to place a reasonable constraint on the initial stellar rotation period. Therefore, we employ the slowly rotating star with a period of 15\,days at the age of 150~Myr, as this maximises the possible atmospheric lifetime.

We note, however, that the majority of the models and empirical approximations describing XUV evolution, including the one used here, describe an average star of a given type. Therefore, besides the dependency on the initial rotation rate, for real stars, one can expect a spread around predicted XUV values within a factor of a few \citep[][]{johnstone2021mors}, also during the initial saturation phase ($\sim$\,37\,Myr for TOI-238\,b) when XUV luminosity depends weakly on rotation. For some planets, the host star being significantly fainter than average for its type can facilitate keeping a hydrogen-dominated atmosphere within the Neptunian desert \citep{Fernandez2024MNRAS.527..911F}. For TOI-238\,b, however, even a further decrease in XUV irradiation by an order of magnitude \citep[which does not linearly increase the atmospheric evaporation time, e.g.][]{kubyshkina_2021mesa} would not lead to a sufficient decrease in mass loss rates for the planet to keep its atmosphere until its present age, as we show below.

\begin{figure}
    \centering
    \includegraphics[width=\hsize]{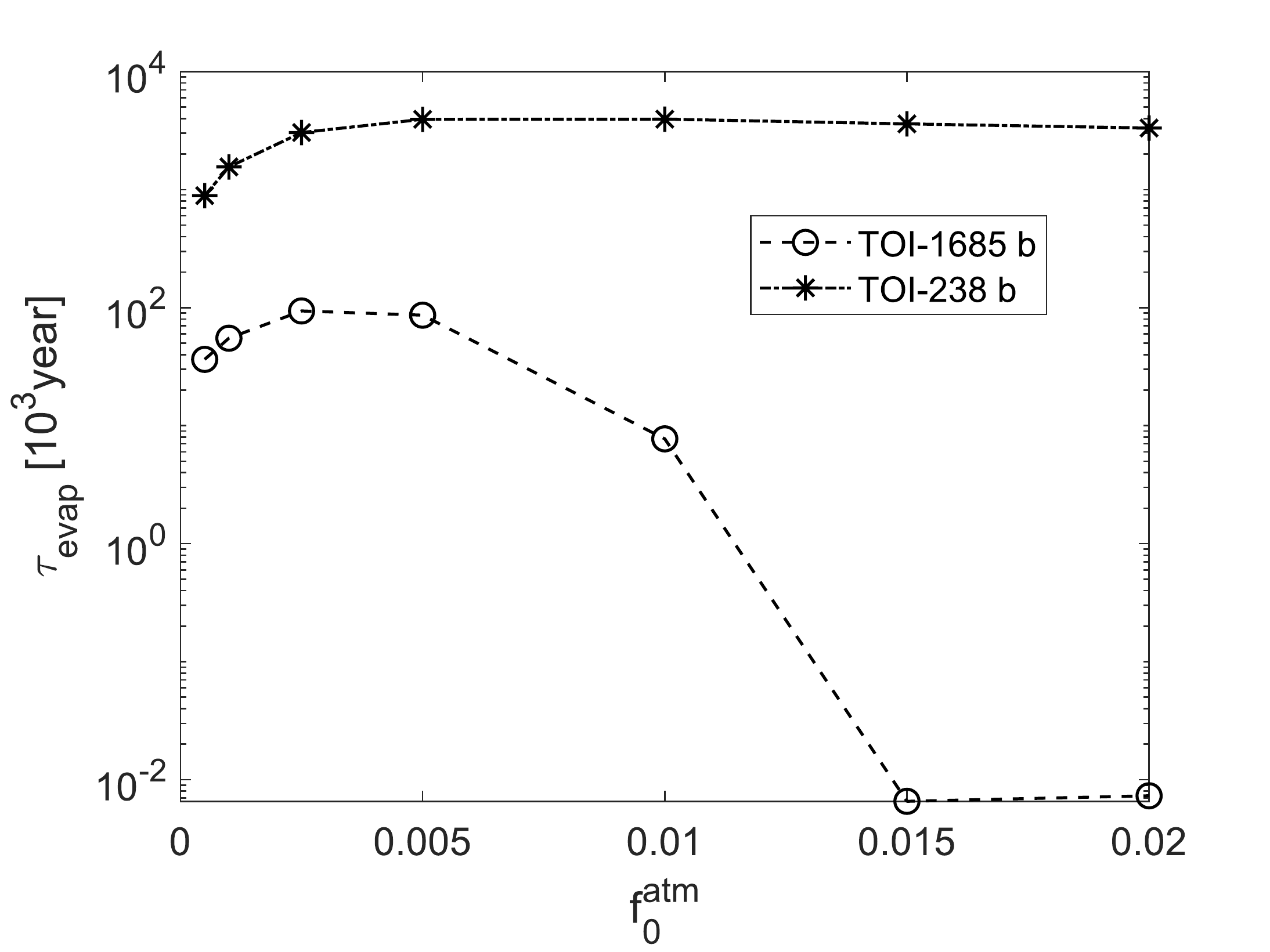}
    \caption{Total evaporation time of the atmospheres of TOI-238\,b (dash-dotted line with asterisks) and TOI-1685\,b (dashed line with circles) as a function of the initial atmospheric mass fractions.}
    \label{fig:HevaporationPlot}
\end{figure}

The evaporation time, $\tau_{\rm evap}$, of the atmosphere of TOI-238\,b can also depend, given the planetary and stellar parameters discussed above, on the initial atmospheric mass (see Figure~\ref{fig:HevaporationPlot}). This parameter can to some extent be constrained by formation models \citep[e.g.][]{Emsenhuber+2021a}, but such an estimate is dependent on the assumptions of the formation scenario and parameters of the protoplanetary disk. Therefore, we remain agnostic towards the formation scenario and consider the full range of initial atmospheric mass fractions, $f_{\rm 0}^{\rm atm}$, between 0.05 and 2\%. Higher atmospheric mass fractions, combined with a high temperature expected for newly born planets, lead to the over-inflation of the planetary radius and the disruption of the atmospheres due to the Roche lobe outflow according to our models.

In Figure\,\ref{fig:HevaporationPlot}, we present the results of our simulations. One can see that the time needed for the evaporation of the atmosphere, $\tau_{\rm evap}$, changes depending on the initial atmospheric mass fraction and maximises at average $f_{\rm 0}^{\rm atm}$ values. This is due to the interplay between the increasing atmospheric mass, which leads to an increase of $\tau_{\rm evap}$ at low $f_{\rm 0}^{\rm atm}$, and the increase of the atmospheric mass loss rates due to the atmospheric inflation. More specifically, mass loss rates depend on planetary radius stronger than to the power of three \citep[e.g.][]{kubyshkina2018_grid}, which compensates for the increase of the atmospheric mass and leads to a decrease of $\tau_{\rm evap}$ at high $f_{\rm 0}^{\rm atm}$. A more detailed discussion of this effect can be found, for example, in \citet{kubyshkina_2021mesa}. 
Thus, for TOI-238\,b, the atmospheric lifetime maximises at $f_{\rm 0}^{\rm atm}$$\sim$0.5\% and varies between $\sim$380 thousand years and $\sim$1.3\,Myr. These results safely rule out the possibility that TOI-238\,b could retain any fraction of their primordial H/He-dominated atmospheres.

We note that the position of the maximum in $\tau_{\rm evap}(f_{\rm 0}^{\rm atm})$ and its value are to some extent dependent on the assumptions of our models and specifically on the initial (core) temperature of the planet, which is poorly constrained. In the present models, the initial temperature increases with increasing $f_{\rm 0}^{\rm atm}$ from $\sim$1300\,K to $\sim$4700\,K, adding up to the inflation of the atmosphere. Fixing this value at $\sim$3000\,K throughout would lead to a shift of the maximum in $\tau_{\rm evap}(f_{\rm 0}^{\rm atm})$ at values of $f_{\rm 0}^{\rm atm}$ about a factor of two larger, and to a reduction of the maximum atmospheric lifetimes by a factor of about 2--4. Instead, reducing the initial temperatures, even to unrealistic values, does not lead to an increase in the atmospheric lifetimes of more than a factor of two. Therefore, our assumed initial temperature does not affect our conclusions.

\subsubsection{Evaporation of water-rich atmospheres}\label{sec:evap_h2o} 
In the previous sections, we have demonstrated that pure H/He atmospheres could not be stable for TOI-238\,b, given the planet's low mass and close-in orbit. 
We have further shown that the present-day parameters of the planet can be well reproduced assuming that its atmosphere consists of water vapour. 
Due to the large mean molecular weight $\mu$ of water ($\mu_{\rm H2O}\simeq 18 \mu_{\rm H}$), these kinds of atmospheres are considerably more compact than hydrogen-dominated ones, meaning that even substantial water atmospheres result in relatively small planetary radii. This is true even for the large core luminosities typical for young planets (see \ref{sec:int_struc_238}). In terms of atmospheric escape, this implies that the interaction area of a planet, more specifically the stellar irradiation absorption area in the case of XUV-driven evaporation, and hence the mass loss rates are much smaller than those of the hydrogen-dominated atmospheres. Combined with the fact that water molecules are known to be an effective cooling agent \citep[e.g.][]{johnstone2020,yoshida2022,GMunoz2023A&A...672A..77G,GMunoz2024Icar..41516080G}, one can expect that water vapour atmospheres are more stable against evaporation.
However, given that we obtained extremely short lifetimes for H/He atmospheres, we will test this possibility more thoroughly.

To date, there are only a few models capable of accurately modelling the escape of water (or water-rich) atmospheres. Moreover, sophisticated models, such as \citet{johnstone2020}, \citet{yoshida2022} or \citet{GMunoz2024Icar..41516080G}, require a computation time that is too high to be applied self-consistently in the frame of planetary atmospheric evolution, with at least thousands of estimates necessary for a single planet across its lifetime. Therefore, to estimate how much of the water atmosphere could be lost throughout the lifetime of TOI-238\,b, we employ a simplified, analytical, energy-limited model predicting the atmospheric mass loss rate for the given planetary parameters as
\begin{equation}\label{eq:EL}
    \Dot{M}_{\rm XUV} = \frac{\pi\eta F_{\rm XUV}R_{\rm pl}R^2_{\rm eff}}{KGM_{\rm pl}}\,,
\end{equation}
where $R_{\rm pl}$ and $M_{\rm pl}$ are the observed transit radius (i.e. where the optical depth $\tau\,\simeq\,2/3$) and the mass of the planet, $F_{\rm XUV}$ is the stellar XUV flux at the planetary orbit, $G$ is the gravitational constant and the factor $K$<1 accounts for the stellar tidal forces \citep{erkaev2007}. The effective radius of XUV absorption $R_{\rm eff}$ for the fixed energy of incoming photons $E_{\rm \lambda}$ can be approximated as \citep[e.g.][]{chen2016ApJ...831..180C}
\begin{equation}\label{eq:Reff}
    R_{\rm eff} = R_{\rm pl} + H\log\left(\frac{P_{\rm photo}}{P_{\rm E\lambda}}\right)\,,
\end{equation}
where $H = (k_{\rm b}T_{\rm eq})/(2\mu g)$ is the atmospheric scale height at the photosphere, with $g = GM_{\rm pl}/R^2_{\rm pl}$. $P_{\rm photo}$ is the photospheric pressure, which can take values between a few tens and a few hundreds of mbar for H/He atmospheres. Here, we employ the 100\,mbar value for all planetary parameters, which has a negligible influence on the results. Finally, $P_{\rm E\lambda} = \mu g/\sigma$ is the pressure at the absorption level of photons, with the absorption cross-section $\sigma = 6\times10^{-18} ({E_{\rm \lambda}}/{13.6 \textrm{eV}})^{-3}\,{\rm cm^2}$. 
We note that this cross-section is for the hydrogen absorption, as we expect that in ${\rm H_2O}$-atmospheres the XUV heating will be led by the photoionisation of H atoms, dissociated from water molecules. To account for the differences in absorption heights of photons with different energies, we split the $F_{\rm XUV}$ in Equation\,\eqref{eq:EL} into X-ray and EUV bands, which we assign the photon energies of 250\,eV and 20\,eV, respectively, following \cite{kubyshkina2018_grid}. We then calculate $R_{\rm eff}$ and the mass loss rates for each band, so that $\Dot{M}_{\rm XUV} = \Dot{M}_{\rm X} + \Dot{M}_{\rm EUV}$. 
Moreover, we performed hydrodynamic simulations of water-rich atmospheres for the present-day parameters of TOI-238\,b, as will be discussed in more detail in Section\,\ref{sec:hydro_toi238}. The analysis of these simulations suggests that despite the high molecular weight of such atmospheres, the XUV heating occurs at altitudes similar to those predicted by Equation\,\eqref{eq:Reff} for pure H/He atmospheres. Therefore, exclusively for calculating $R_{\rm eff}$, we employ $\mu\simeq 1.3\,m_{\rm H}$.

Finally, the term $\eta$ in Equation\,\eqref{eq:EL} is the so-called heating efficiency coefficient, which parametrises the total fraction of the incoming XUV irradiation spent specifically on heating the atmosphere. We note that realistically, the heating efficiency is not constant across the specific atmosphere and depends on the altitude, photon energy, and the composition of the atmosphere \citep[e.g.,][]{shematovich2014,johnstone2018,kubyshkina2024A&A...684A..26K}. The parameter $\eta$, in turn, is meant to fit the mass loss rates predicted by Equation\,\ref{eq:EL} to the predictions of more sophisticated models \citep[e.g.][]{salz2016,caldiroli2022} or to observationally-based predictions \citep[e.g.][]{Owen+Wu2017}. For hydrogen-dominated atmospheres, the value of $\eta$ varies between 0.1-0.3, while the heating efficiency in water-rich atmospheres is expected to be significantly lower. The existing estimates of $\eta$ remain quite sparse and typically bound to the specific planet type in question. For an Earth-mass planet with a ${\rm H_2O}$ atmosphere orbiting a Sun-like star at 1\,AU, \citet{johnstone2020} estimates the value of $\eta$ as low as 0.01. Meanwhile, for mixed ${\rm H_2-H_2O}$ atmospheres around terrestrial planets in the habitable zone of M-dwarfs, \citet{yoshida2022} predict a decrease in the heating efficiency compared to pure hydrogen atmospheres of at least a few times. However, applying these results directly to model hot Neptune planets would be poorly justified. Therefore, in our simulations, we consider the whole range of $\eta$ between 0.01--0.15, with the prediction of \citet{johnstone2020} as a minimum and the value typically used for H/He atmospheres \citep{salz2016} as a maximum. 
\begin{figure}
    \centering
    \includegraphics[width=\linewidth]{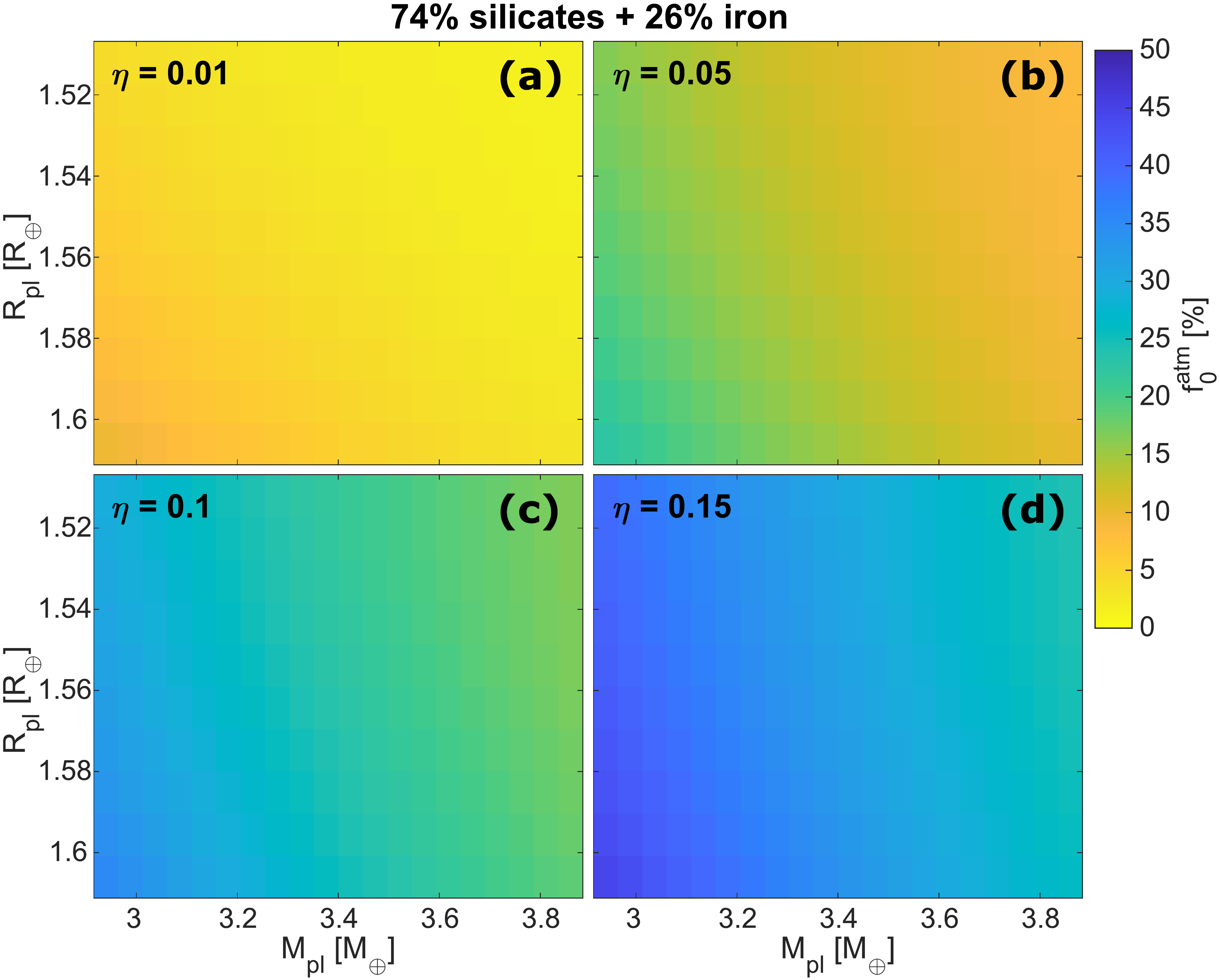}
    \caption{The primordial atmospheric (water vapour) mass fraction of TOI-238\,b against present-day mass and radius of the planet, as given by the observational constraints. Four panels correspond to different values of the heating efficiency parameter $\eta$: 0.01 (a), 0.05 (b), 0.1 (c), and 0.15 (d). In all four cases, the planetary core has an Earth-like composition and the luminosity of $10^{21}$~${\rm erg/s/cm^2}$, and the star evolved as a slow rotator.}
    \label{fig:H2Oevap2D-TOI238b}
\end{figure}

To estimate the total amount of water that the planet could lose throughout its lifetime given the known present-day parameters, we invert the evolution algorithm used to study the evaporation of H/He atmospheres. We do this for different values of $\eta$, different core compositions (pure rock, Earth-like, and Mercury-like), intrinsic planetary luminosities ($10^{19}$, $10^{21}$, and $10^{23}$\,${\rm erg/s}$), present-day age estimates between 2.7 and 12.4\,Gyr \citep{Mascareno+2024}, and combinations of $M_{\rm pl}$ and $R_{\rm pl}$ within the $1\sigma$ interval given in section\,\ref{sec:obs_238}. For a specific set of parameters, we start by estimating the present-day mass of a pure ${\rm H_2O}$ atmosphere for the given mass and radius values of the planet by interpolating between the mass-radius relations that the interior structure inference in Section\,\ref{sec:int_struc_238} is based on. We then define the atmospheric mass loss rate using Equation\,\eqref{eq:EL}, for the present-day planetary parameters and stellar XUV flux, as well as the chosen $\eta$ value. As we are particularly interested in the maximum amount of water that can be evaporated, for the stellar model we employ the predictions of the \texttt{Mors} code \citep{johnstone2021mors} for the same stellar mass as for the analysis above for H/He atmospheres, but assuming the fast rotator scenario ($P_{\rm rot} = 1$\,day at 150\,Myr) along with the slow rotator. This gives us the XUV flux at a given time. After defining the present-day mass loss rate, we set a time step in a way that no more than 0.5\% of the total water mass can be lost within this time frame. We then step back in time, adjusting the mass of ${\rm H_2O}$ (hence, the water mass fraction and the mass of the planet) and re-defining the planet's radius using the same mass-radius relations. We repeat this procedure until we reach an age of 10\,Myr (the adopted protoplanetary disk dispersal time) or a water mass fraction of 50\%. In the latter case, we assume that the water atmosphere could have fully evaporated for the given present-day parameters, as formation models predict that the primordial water fraction is unlikely to be higher than 50\%.

In addition to the mass loss rate predicted by Equation\,\eqref{eq:EL}, we estimated the core-powered mass loss rates adjusted for water molecules. These are controlled by the bolometric heating from the host star \citep{gupta_schlichting2019MNRAS.487...24G}. We found, however, that they are negligible compared to the XUV-driven atmospheric mass loss, even though in the case of H/He atmospheres, core-powered escape is comparable or even dominant relative to XUV-driven escape at young ages under the conditions of TOI-238\,b (depending on other assumptions of the model). As a result, we obtain estimates of the initial water mass fractions and planetary masses at the age of 10\,Myr, in dependence on the present-day mass and radius of the planet for different model assumptions (core types, ages, and heating efficiencies).
\begin{figure}
    \centering
    \includegraphics[width=\linewidth]{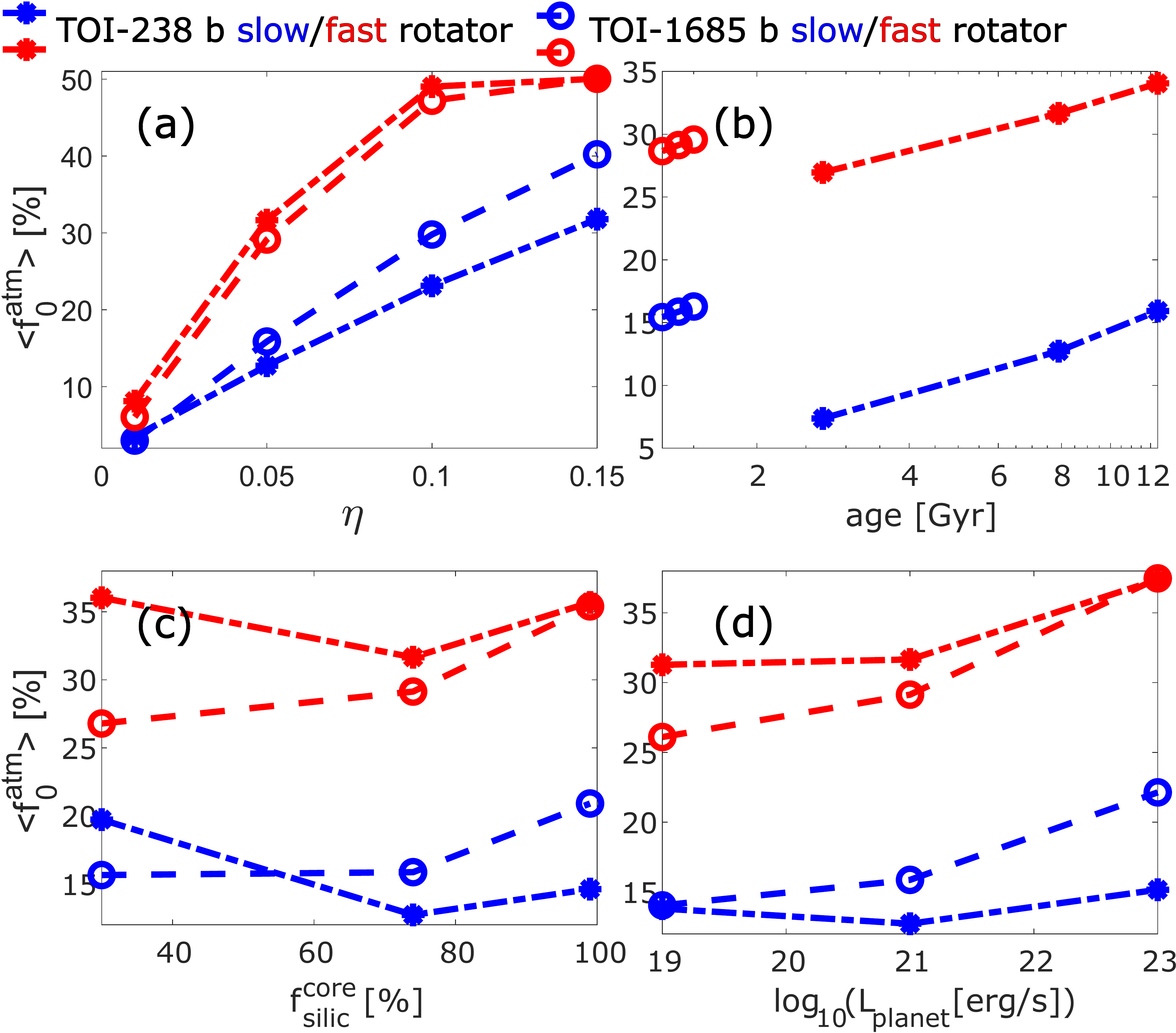}
    \caption{Dependence of the average initial mass fraction $<f_{0}^{\rm atm}>$ of TOI-238\,b (dash-dotted lines with asterisks) and TOI-1685\,b (dashed lines with circles) on the adopted heating efficiency parameter $\eta$ (panel a), present age of the system (b), composition of the core (namely, silicates fraction; panel c), and core luminosity (d). Different colours show the different stellar rotation scenarios, as indicated in the legend. The age intervals in panel (b) were chosen according to the age estimates of the two systems as derived in Sections\,\ref{sec:starTOI238} and \ref{sec:starTOI1685}.} 
    \label{fig:H20evap-TOI238b_pdep}
\end{figure}

In Figure\,\ref{fig:H2Oevap2D-TOI238b}, we show our predictions for the initial water atmosphere mass fraction $f_{\rm 0,H2O}^{\rm atm}$ of TOI-238\,b in dependence on its present-day mass and radius estimates within 1$\sigma$ and heating efficiency parameter $\eta$. Here, we assumed that the planetary core has an Earth-like composition and an intrinsic luminosity of $10^{21}$~${\rm erg/s}$, the star evolved as a slow rotator ($P_{\rm rot}^{\rm 150 Myr}\,=\,15$\,days), and the present age of the system is 7.9\,Gyr. In all cases, $f_{\rm 0}^{\rm atm}$ maximises for lower masses and larger radii, as this combination maximises atmospheric photoevaporation due to lower gravity and a larger photoionisation heating area. At the same time, the atmospheric photoevaporation minimises at higher masses and smaller radii. Thus, in each case the possible values of $f_{\rm 0}^{\rm atm}$ vary over a wide range, changing from $\sim$1.5-10\,\% if $\eta\,=\,0.01$ to $\sim$27-44\,\% if $\eta\,=\,0.15$. Overall, accounting for all possible combinations of planetary interior and stellar parameters, the possible value of $f_{\rm 0}^{\rm atm}$ vary between $\sim$0.03\,\% and $\sim$50\,\%. 

In Figure\,\ref{fig:H20evap-TOI238b_pdep}, we illustrate the dependence of the average $f_{\rm 0}^{\rm atm}$ (corresponding to the middle values of $R_{\rm pl}$\,=\,1.559\,R$_{\oplus}$ and $M_{\rm pl}$\,=\,3.4\,M$_{\oplus}$ in Figure\,\ref{fig:H2Oevap2D-TOI238b}) on the assumed age of the system, the composition and luminosity of the core, and the parameter $\eta$. In each of the cases, we vary one of these four parameters, while the other three are kept at ``average'' values, meaning $L_{\rm pl}$\,=\,10$^{21}$\,${\rm erg/s}$, Earth-like core, age of the system of 4.3\,Gyr and $\eta$ = 0.05. One can see that the estimate depends strongly on the adopted value of $\eta$ with changes of a factor of 5-10 across the considered interval. The value also changes by a factor of 2 with the age of the system. The dependence on the composition and luminosity of the core is weaker and less pronounced, due to the complicated interplay between atmospheric structure and evaporation. On average, one can expect stronger evaporation and therefore a larger $f_0^{\rm atm}$ for more silicate-rich and hot cores.

\subsubsection{Hydrodynamic simulations}\label{sec:hydro_toi238}
Finally, to have a better understanding of the atmospheric heating efficiency that one can expect for the hot sub-Neptune TOI-238\,b in the case of a water-rich atmosphere, we performed a series of hydrodynamic simulations employing the Cloudy e Hydro Ancora INsieme code \citep[CHAIN;][]{kubyshkina2024}. We adopted the median values of the present-day parameters of the planet, namely $R_{\rm pl}$\,=\,1.559\,R$_{\oplus}$, $M_{\rm pl}$\,=\,3.4\,M$_{\oplus}$, and $a$\,=\,0.02556\,AU. We further assumed that the age of the system is $\sim$2.7\,Gyr, thereby assuming that the XUV irradiation is still high with values around $\sim$10$^4$\,${\rm erg/s/cm^2}$.
Following the procedure introduced in \cite{Egger+2024}, we assumed that the water was accreted in ice form and was later turned to steam mixed with the hydrogen-helium envelope as the planet migrated inwards. 
We control the water mass fraction in the atmosphere by setting the abundance of the atomic oxygen relative to hydrogen. In this approach, the actual number densities of water molecules in the upper atmospheres are decided by the photochemistry solver.

\begin{figure}
    \centering
    \includegraphics[width=1\linewidth]{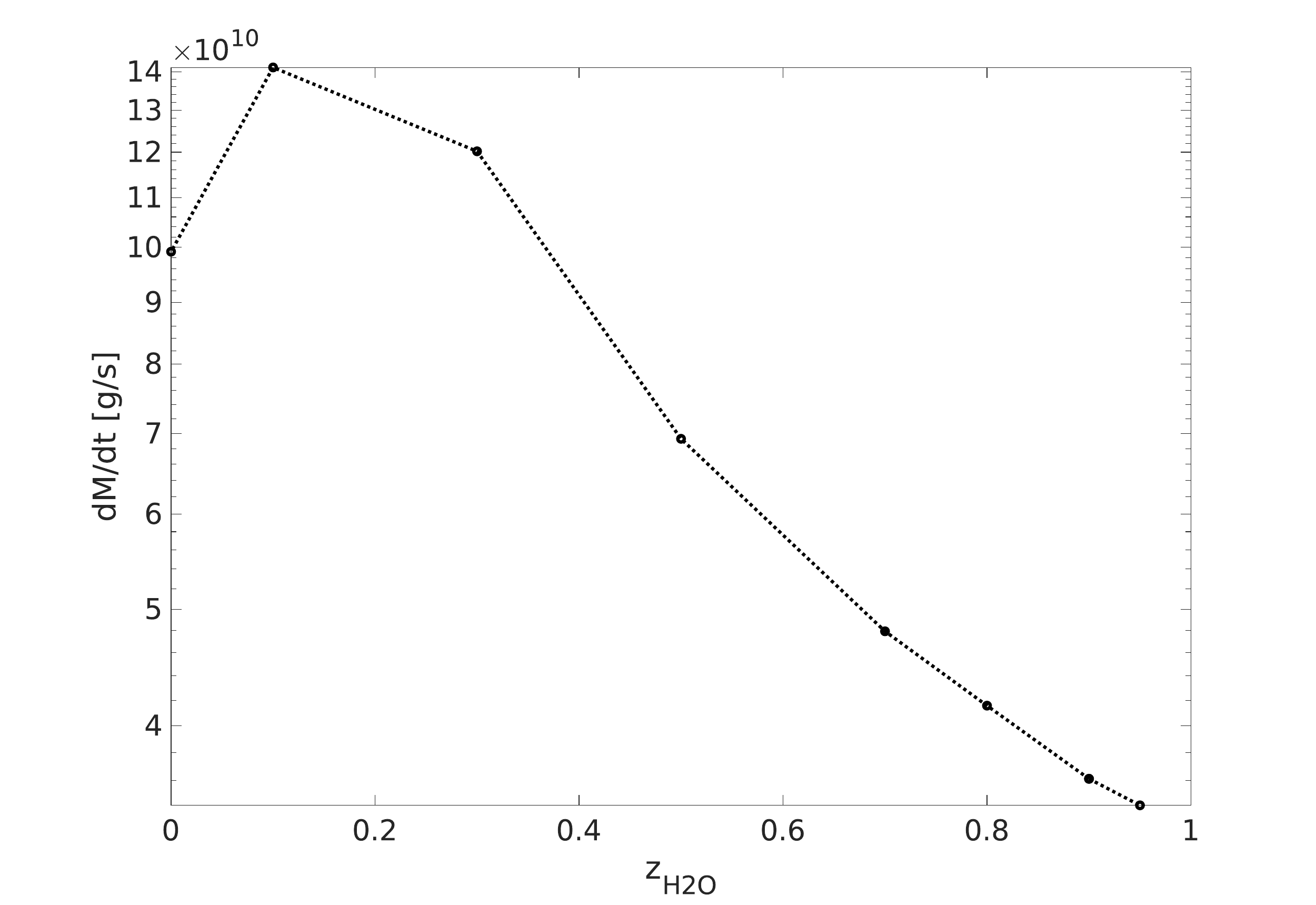}
    \caption{Atmospheric mass loss rates of TOI-238\,b in dependence on the water mass fraction in the atmosphere.}
    \label{fig:evap_H2O_hydro}
\end{figure}

We considered water mass fractions $Z$ of 10\%, 30\%, 50\%, 70\%, 80\%, 90\%, and 95\%, where the lowest mass fraction corresponds to an oxygen enrichment of $\sim$13 times over solar abundance and the highest one of $\sim$700 times. We adjusted the mean molecular weight consistently, and compared our results to pure H/He atmospheres with a Sun-like metallicity. Our results for the atmospheric mass loss rates are shown in Figure\,\ref{fig:evap_H2O_hydro}. One can see that at low water enrichment (10\% and 30\%), the atmospheric evaporation is higher than for a pure H/He atmosphere. This happens because the number of water molecules that can survive in the regions relevant for atmospheric escape is low given the harsh irradiation environment of TOI-238\,b. This then means that their contribution to the cooling and the changes of the mean molecular weight of the atmosphere can also be considered negligible. At the same time, the relative abundance of atomic oxygen, photodissociated from water molecules, increases, which contributes to the heating of the atmosphere. With further increase of the water mass fraction, the mass loss rates decline, and at the maximum considered mass fraction of 95\% reach a value 3 times smaller than the pure H/He case. This implies that the heating efficiency parameter needed to reproduce these escape rates with the energy-limited approximation would also be about one-third of that for H/He atmospheres. The latter was estimated to be about $\eta\,=\,0.15$ for similar hydrodynamic models \citep{salz2016}. Therefore, we consider the $\eta$\,=\,0.05 case in the evolution analysis presented above to be most likely.

\begin{figure}
    \centering
    \includegraphics[width=1\linewidth]{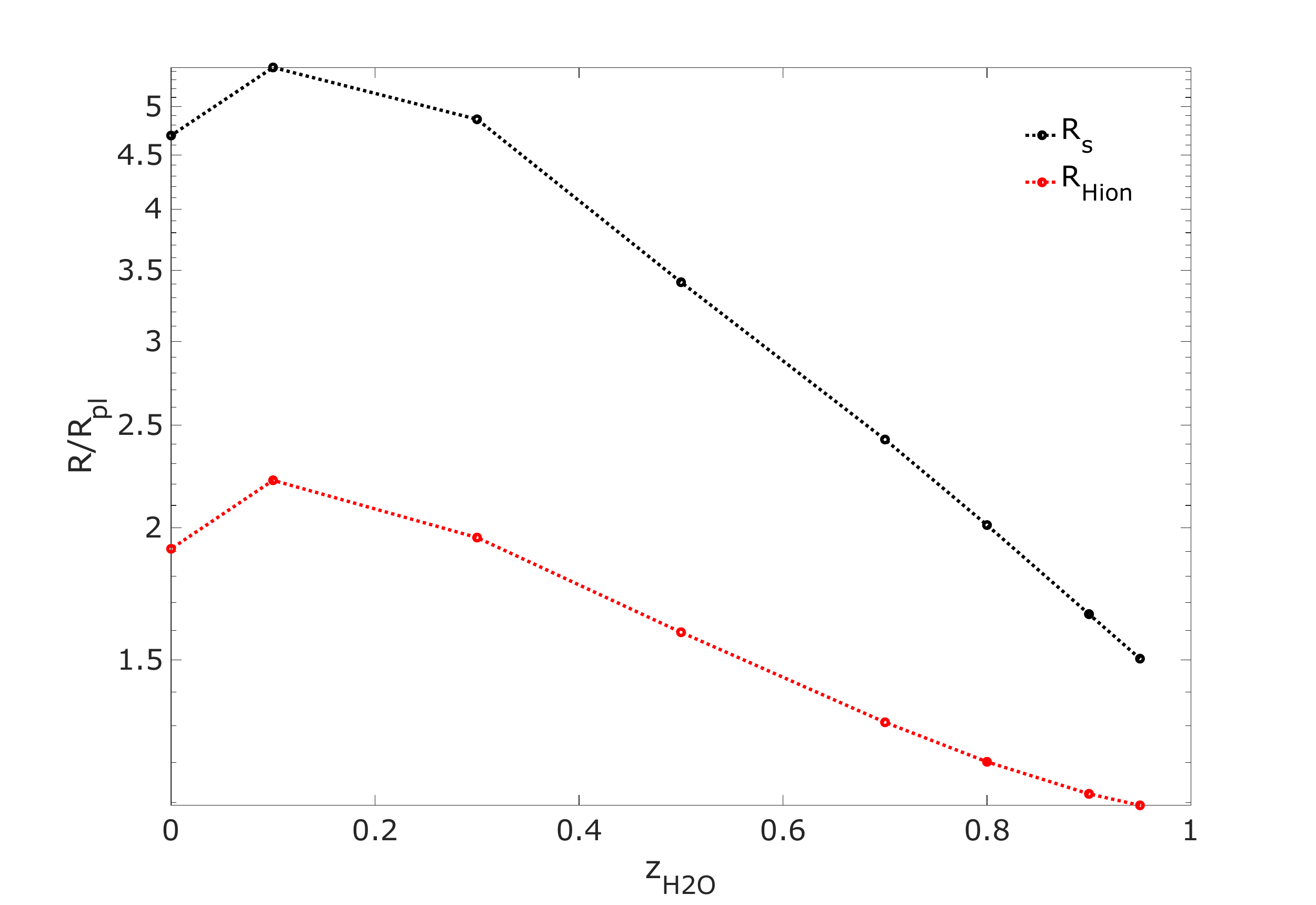}
    \caption{Radial location R of the sonic radius R$_\mathrm{S}$ and ionisation front R$_\mathrm{Hion}$ (maximum of the $\rm H^+$ number density) in the upper atmosphere of TOI-238\,b in dependence on the water mass fraction in the atmosphere.}
    \label{fig:evap_H2O_hydro_radii}
\end{figure}

The second uncertain parameter in Equation\,\eqref{eq:EL} is the effective radius of XUV absorption $R_{\rm eff}$. Assuming a pure H$_2$O atmosphere and therefore $\mu$\,=\,18~$\mu_H$ in Equation\,\eqref{eq:Reff} leads to a very compact $R_{\rm eff}$. In hydrodynamic simulations, at high Z values, this prediction is similar to the position of the peak in the ${\rm H^+}$ number density $n_{\rm H^+}$. However, both in the real case and in our simulations, the XUV spectrum is not limited to one photon energy. Photon absorption therefore occurs at different altitudes, the ion density remains high enough also above the $n_{\rm H^+}$ peak, and the photoionisation heating can remain significant over a wide range of altitudes. The high temperature and low gravity of TOI-238\,b facilitate the expansion of the upper atmosphere (i.e., stretching of characteristic distances) and the heating is significant up to the sonic point, which is $R_{\rm s}\simeq1.5\,R_{\rm pl}$ for $Z\,=\,0.95$ (see Figure\,\ref{fig:evap_H2O_hydro_radii}).

We used the results of these hydrodynamic simulations to optimise the input parameters of the evolution models described in Section\,\ref{sec:evap_h2o}. With this optimisation, the present-day mass loss rates that we obtain with the simplified formulation given by Equation\,\eqref{eq:EL} and by assuming $\eta = 0.05$ are similar to those obtained using the more sophisticated model. 
For a further evaluation of the results and to test whether the uncertain parameters, in particular the heating efficiency, vary throughout the evolution would require a self-consistent implementation of the hydrodynamic model within the evolutionary framework. This is currently not possible due to the high computational costs.

\subsection{TOI-1685 b}
\label{sec:TOI-1685b}

TOI-1685~b is a transiting ultra-short-period planet orbiting an M3V star with an orbital period of 0.669 days. The planet is marked in red in Figure~\ref{fig:demographics} and seems to lie clearly inside the corresponding Hot Water World triangle. However, similarly to TOI-561~b, multiple different sets of mass and radius values exist in the literature for this planet, as is shown in the middle panel of Figure~\ref{fig:triangle_obs}. TOI-1685~b was first discovered as a transiting planet candidate by TESS in sector 19 and then verified by two independent groups using both additional ground-based photometry and radial velocity observations from CARMENES and IRD respectively. While \cite{Bluhm+2021} find mass and radius values corresponding to a bulk density of only $\rho = 4.21^{+0.95}_{-0.82}$~g~cm$^{-3}$, \cite{Hirano+2021} report the planet to be most likely rocky with a bulk density of $\rho = 6.1^{+1.9}_{-1.6}$~g~cm$^{-3}$. While the two mass values were derived from two different data sets, they are consistent within the uncertainties, while the radius values show a significant discrepancy, which is caused by different values for both the stellar radius as well as the transit depth. 

Since then, two more works have investigated the characteristics of TOI-1685~b. \cite{Luque+Palle2022} performed a joint analysis of all radial velocity and transit observations of the two previous works and find a radius identical to \cite{Bluhm+2021} but with a slightly smaller mass, while \cite{Burt+2024} include an additional sector of TESS photometry, new ground based photometry and additional radial velocity data and find a radius that is, within the error bars, compatible with the one of \cite{Hirano+2021}, as well as a mass compatible with \cite{Luque+Palle2022}. This is again partly due to differences in the derived stellar parameters as well. TOI-1685~b is also the target of three cycle~2 JWST programs \citep{Luque+2023JWST,Benneke+2023JWST,Fisher+2023JWST}. Both \cite{Bluhm+2021} and \cite{Hirano+2021} suggested that there might be an additional planet in the system, at orbital periods of 9~days and 2.6~days respectively. \cite{Burt+2024} tested the planetary nature of both of these signals, but find the 9~day signal likely to be linked to the star's rotational modulation and the 2.6~day signal due to systematics of the IRD data used by \cite{Hirano+2021}.

\subsubsection{Host star characterisation}\label{sec:starTOI1685} 

The spectroscopic parameters for TOI-1685, so the effective temperature and the metallicity, were derived using the latest ODUSSEAS code \citep[][]{Antoniadis-Karnavas+2024}. For this we used the template CARMENES spectrum for this star that was taken from the spectral library provided in \citet[][]{Nagel+2023}. The trigonometric surface gravity for this star was derived using recent Gaia data following the same procedure as described in \citet[][]{Sousa+2021}.
While our derived value for the effective temperature is in agreement with the ones found by \cite{Bluhm+2021}, \cite{Hirano+2021}, and \cite{Burt+2024}, the derived values for the metallicities vary drastically, from $-0.13\pm0.16$ \citep{Bluhm+2021} over our own value of $0.01\pm0.12$ up to $0.14\pm0.12$ \citep{Hirano+2021} and $0.3\pm0.1$ \citep{Burt+2024}.

Determining the stellar radii of M-dwarf stars using SED-based methods is notoriously challenging due to the unknown inaccuracies of M-dwarf stellar atmospheric models. Therefore, we utilised an empirically-derived absolute magnitude-metallicity-radius calibration \citep{Mann2015}. We combined the published $K_{\rm s}$ magnitude (5.882$\pm$0.020) with our derived metallicity to produce our stellar radius listed in Table~\ref{tab:stellar_params} with the uncertainty inflated by the empirical calibration uncertainty \citep{Mann2015}. This value is in agreement with the ones found by \cite{Hirano+2021} and \cite{Burt+2024}, but significantly smaller than the value reported by \cite{Bluhm+2021}.

We followed a strategy similar to that described in Section~\ref{sec:starTOI238} to derive the mass of TOI-1685, hence obtaining $M_{\star}=0.466\pm0.019\,M_{\odot}$. On the one hand, given the slow evolution of M dwarfs along the HR diagram, an isochrone-based estimate of the stellar age is not meaningful. On the other hand, \citet{engle2023} recently developed gyrochronological relations specifically set up for M dwarfs to assess their evolutionary state. Plugging in the stellar rotation period $P_{\mathrm{rot}}=18.2\pm0.5$\,d \citep{Burt+2024}, we computed a gyro-age of $t_{\star,\mathrm{gyro}}=1.4\pm0.1$\,Gyr. According to \citet{kiman2021}, we note that TOI-1685 is classified as inactive, since its $H_{\alpha}\text{-EW}=0.51\pm0.06$\,\AA\ \citep{Bluhm+2021} is lower than the $G-G_{\mathrm{RP}}$ colour-dependent threshold  given by $H_{\alpha}\text{-EW}_{\mathrm{bound}}=0.82$\,\AA\ \citep[$G-G_{\mathrm{RP}}=1.14$;][]{GaiaDR3-2023}. An inactive M3 dwarf is likely older than $\sim$\,1\,Gyr \citep[see Fig.~5a in][]{kiman2021}, which is in agreement with our gyrocronological estimate. Our derived value for the stellar mass is in agreement with the ones found by \cite{Hirano+2021} and \cite{Burt+2024} at the 1$\sigma$ level and with the one reported by \cite{Bluhm+2021} at the 2$\sigma$ level. Our value for the stellar age agrees within 2$\sigma$ with the estimate from \cite{Burt+2024}.

Due to significant line blending, determining the individual elemental abundances of M-dwarfs from visible spectra is challenging \citep[e.g.,][]{Maldonado-20}. We estimated the abundances of Mg and Si by following the methodology outlined in \citet{Demangeon-21}. We used the radial velocity, parallax, coordinates, and proper motions from \textit{Gaia} DR3 to derive the Galactic space velocities (UVW) of TOI-1685 via the \texttt{GalVel\_Pop.py}\footnote{\url{https://github.com/vadibekyan/GalVel_Pop/blob/main/GalVel_Pop.py}}. We derived U = $-$24.9 ± 0.4 km/s, V = $-$17.7 ± 0.1 km/s, and W = 4.1 ± 0.1 km/s relative to the local standard of rest, adopting the solar peculiar motion from \citet{Schonrich-10}. Based on these velocities, and using the characteristic parameters of Galactic stellar populations from \citet{Reddy-06}, in conjunction with the approach of \citet{Adibekyan-12b}, we estimated a 99\% probability that TOI-1685 belongs to the Galactic thin disk. Additionally, from the APOGEE DR17 data \citep{Abdurrouf-22}, we selected cool stars with metallicities similar to TOI-1685 and classified as part of the chemically defined Galactic thin disk. This yielded a sample of several thousand stars, for which we calculated the mean Mg and Si abundances, along with their standard deviations (representing star-to-star scatter). We found [Mg/H] = 0.04 ± 0.17 dex and [Si/H] = 0.02 ± 0.17 dex, as listed in Table~\ref{tab:stellar_params}.

\subsubsection{Observations and data analysis} 
\begin{figure*}
    \centering
    \includegraphics[width=\textwidth]{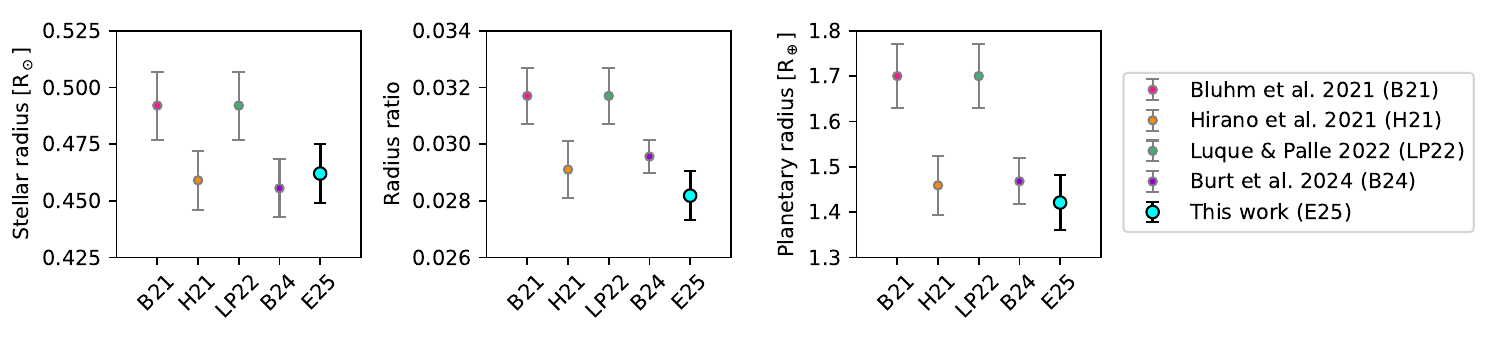}
    \caption{Comparison of the different values in the literature for the stellar radius, radius ratio and planetary radius of TOI-1685~b.}
    \label{fig:radius_comp_1685}
\end{figure*}

TOI-1685 was observed by TESS in sectors~19 (December 2019 at two-minute cadence) and 59 (December 2022 at 20-second cadence). Additionally, we observed TOI-1685 for a total of 124.6~hours with CHEOPS as part of the GTO, split over 16~visits between October 2023 and January 2024. The undetrended CHEOPS light curves are shown in Figure~\ref{fig:undetrended_CHEOPS_1685}. Additionally, all CHEOPS observations are summarised in Table~\ref{tab:file_keys}. We again used the publicly available code \texttt{chexoplanet} to perform a joint fit of all available TESS and CHEOPS photometry, as described in more detail in Section~\ref{sec:obs_238}. The ground-based observations used in all four previous models in the literature \citep{Bluhm+2021,Hirano+2021,Luque+Palle2022,Burt+2024} appeared to show discrepancies in radius ratio, potentially due to unmodelled correlated noise and/or impact from atmospheric seeing. In order to provide an independent measurement, we used only space-based observations from TESS and CHEOPS.

We find a value of 0.02818$\pm$0.00086 for the radius ratio, which gives us a planetary radius of 1.421$\pm$0.060~R$_\oplus$ together with our derived stellar radius. This is in agreement with the values found by \cite{Hirano+2021} and \cite{Burt+2024} within one sigma, but is significantly smaller than the radii derived by \cite{Bluhm+2021} and \cite{Luque+Palle2022}. This is due to differences in both the stellar radius as well as the radius ratio, as is visualised in Figure~\ref{fig:radius_comp_1685}. Furthermore, we combined the RV semi-amplitude reported by \cite{Burt+2024} with our newly derived stellar mass listed in Table~\ref{tab:stellar_params} to a value for the mass of TOI-1685~b. The posteriors for the planetary parameters can be found in the right column of Table~\ref{tab:planetary_params}, the resulting photometry fits in the right panels of Figure~\ref{fig:phasefolded_transits}.

\subsubsection{Internal structure analysis} 
\label{sec:int_struc_1685}

\begin{figure*}
    \centering
    \includegraphics[width=\textwidth]{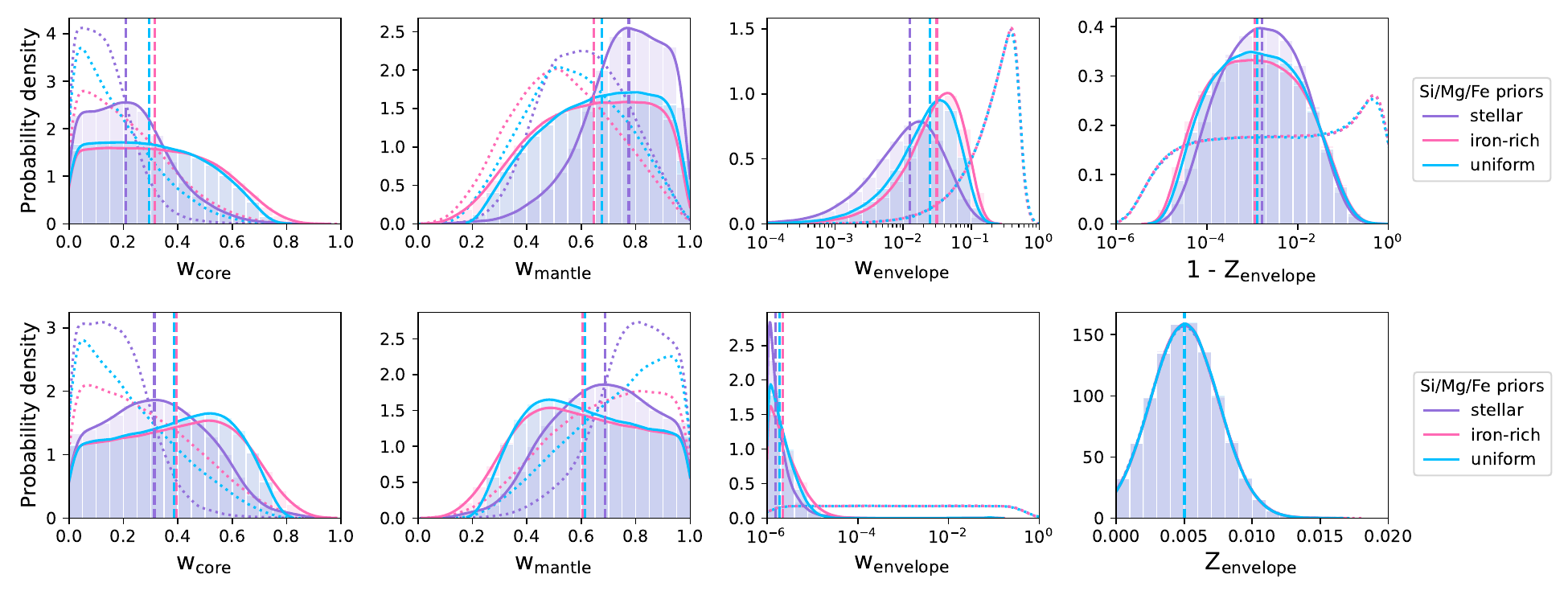}
    \caption{Same as Figure~\ref{fig:int_struct_238} but for TOI-1685~b.}
    \label{fig:int_struct_1685}
\end{figure*}

Analogous to our analysis for TOI-238~b, we now investigate the interior structure of TOI-1685~b. Looking at the mass-radius diagram in the bottom panel of Figure~\ref{fig:triangle_obs}, TOI-1685~b now lies, within 1$\sigma$ in both mass and radius, between the models for a bare core with an Earth-like composition and an iron-less bare core. However, this does not per se mean that the planet does not have a volatile layer and is a bare core, as there are an infinitely large number of possible compositional models that are simply not plotted in Figure~\ref{fig:triangle_obs}. Nonetheless, it does mean that we find TOI-1685~b to lie outside the Hot Water World triangle corresponding to its equilibrium temperature, and therefore cannot conclude that the planet has to contain volatiles, as was the case for TOI-238~b.

Following the same procedure as outlined in Section~\ref{sec:int_struc_238} for TOI-238~b, we then also applied the \texttt{plaNETic} framework to TOI-1685~b, using the stellar and planetary parameters listed in Tables~\ref{tab:stellar_params} and~\ref{tab:planetary_params} as input. The resulting posterior distributions for the most important internal structure parameters are shown in Figure~\ref{fig:int_struct_1685}, while Table~\ref{tab:internal_structure_results_1685} shows the medians and one sigma errors of the posteriors of a wider set of parameters. The results are similar to the ones for TOI-238~b, with volatile layers of a few percent in planetary mass and almost entirely made up of water for a prior inspired by a formation scenario outside the iceline (top row of Figure~\ref{fig:int_struct_1685}), and almost pure H/He envelopes with mass fractions of the order of 10$^{-6}$ for a water-poor prior (bottom row). The main difference is that the inferred water-rich envelopes in the first case are generally slightly smaller and less well constrained for TOI-1685~b compared to the ones for TOI-238~b. For the intrinsic luminosity of TOI-1685~b, we find $\log_{10}\left(\frac{L}{1\textrm{erg s}^{-1}}\right) = 21.34\pm0.08$ based on the stellar age and the planet's inferred internal structure, using the age-luminosity fit of \cite{Mordasini2020}.

\subsubsection{Evaporation analysis} 
\label{sec:evaporation_1685}
To test the stability of H/He and water-dominated atmospheres under the conditions of TOI-1685\,b, we repeat the analysis presented in Sections\,\ref{sec:evap_H/He} and \ref{sec:evap_h2o}. We obtain results that are qualitatively similar to the case of TOI-238\,b. 

Given the relatively young age of the TOI-1685 system, we can use the stellar rotation period estimate of $18.66^{+0.71}_{-0.56}$\,days \citep{Bluhm+2021} to constrain the slowest possible initial rotation rate that still allows to reproduce the present-day parameters of the star. According to the stellar XUV evolution code \texttt{Mors} used in our simulations for the stellar input, this corresponds to a stellar rotation period of 9.6\,days at the age of 150\,Myr.
Thus, we find that a H/He atmosphere would also not be stable at TOI-1685\,b, with even shorter lifetimes than for TOI-238\,b, not exceeding $\sim$100'000 years. The atmospheric lifetimes maximises at $f_0^{\rm atm}\,\simeq\,0.25\%$ and the atmosphere is subject to Roche lobe disruption starting from $f_0^{\rm atm}$\,$\sim$\,1.5\%.

In the case of water-rich atmospheres, our analysis predicts that TOI-1685\,b had to start its evolution with $f_0^{\rm atm}$ similar to those of TOI-238\,b. As in the previous case, even the ${\rm H_2O}$ atmosphere can be removed if one assumes that the star evolved as a fast rotator and the heating efficiency parameter is $\eta\geq0.1$. Concerning the core composition and luminosity, the dependence is more pronounced than for TOI-238\,b, with larger initial atmospheres corresponding to larger internal luminosities and more silicate-rich cores (see Figure\,\ref{fig:H20evap-TOI238b_pdep}), though the differences between different cores remain within $\sim$30\%. Finally, the more narrow age estimate compared to the case of TOI-238\,b diminishes the dependence on the assumed present-day age.
\section{Discussion}
\label{sec:discussion}

\subsection{Model limitations}
\label{sec:discussion_limitations}
\begin{figure}
    \centering
    \includegraphics[width=\linewidth]{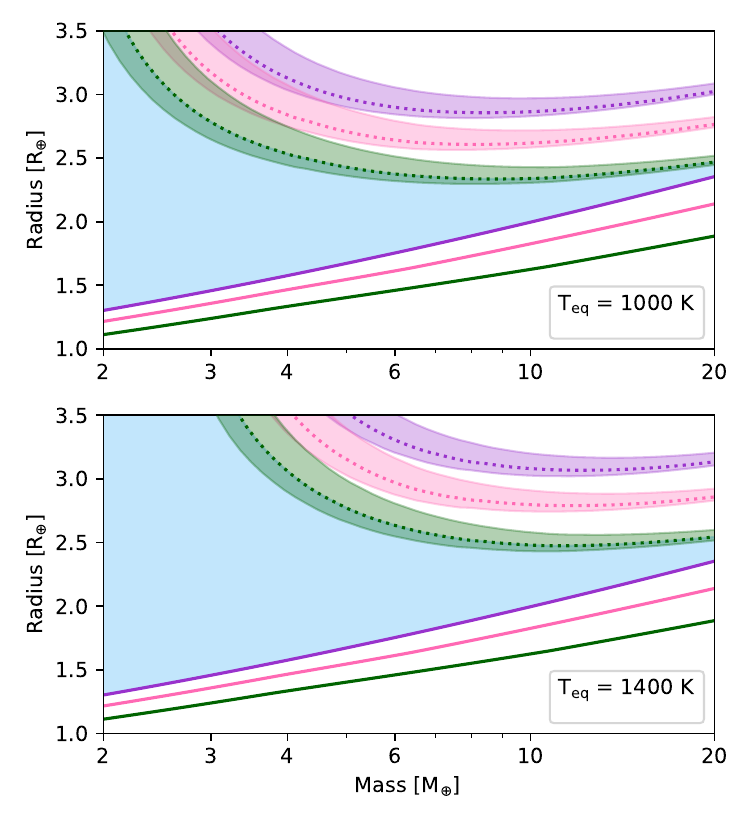}
    \caption{Influence of choosing different values for the intrinsic luminosity of the planet on the upper boundary of the Hot Water World triangles, for a planetary equilibrium temperature of 1000~K (top, similar to the conditions of TOI-1685~b) and 1400~K (bottom, similar to the conditions of TOI-238~b). The dotted lines show the mass-radius models generated using a value of $10^{21}$~erg/s, while the shaded regions visualise the influence of choosing a lower ($10^{20}$~erg/s for the lower boundary) or higher value ($10^{22}$~erg/s for the upper boundary). The remaining line styles and colours are the same as in Figure~\ref{fig:triangle_definition}.}
    \label{fig:triangle_luminosity}
\end{figure}

The concept of Hot Water World triangles introduced in Section~\ref{sec:HWW} allows us to easily decide whether a planet is a potential close-in water world candidate. It therefore gives us an efficient criterion to apply to a large database of exoplanets, without having to run extensive models for each planet individually. However, it is important to note that this concept is based on several simplifying assumptions, which need to be considered in more detail when studying a specific candidate.

One such simplification is the assumption that H/He envelopes with mass fractions of less than 1\% are evaporated on timescales of the order of 1\,Myr, as derived by \cite{Owen+Wu2017} using a minimal analytical model. While we used this approximate value of a 1\% H/He layer to define the upper boundaries of our Hot Water World triangles, for a specific system the exact value of course will always depend on the specific stellar and planetary properties. For some planets, also H/He envelopes that are far more massive than $\sim$1\% would not be expected to survive over long timescales, while for others, also less massive H/He envelopes could potentially survive. There are also many other evolution frameworks that could be used instead, with \cite{Lopez2017}, \cite{Jin+Mordasini2018}, and \cite{Mordasini2020} as examples. \cite{Jin+Mordasini2018} actually also define a `triangle of evaporation' but in the radius-orbital distance space, which contains planets that have lost all their H/He. \cite{Mordasini2020}, on the other hand, gives an approximation for the minimum mass that a planet must have to keep a H/He atmosphere depending on its orbital distance and stellar properties. However, this simplifying assumption of a 1\% H/He layer was only used to select potentially interesting targets for the  observational programme, while we later ran more complex evaporation models in Section~\ref{sec:targets} to calculate whether pure H/He or higher mean molecular weight water envelopes would be expected to survive over long timescales for the two planets we observed here. This will especially become important when observing a planet near the upper boundary of the Hot Water World triangle.

It is also important to consider that mass-radius models for H/He-dominated atmospheres depend on the planet's intrinsic luminosity. This implies that the entropy changes due to each planet’s thermal evolution and atmospheric evaporation should be taken into account \citep[e.g.][]{Rogers+2023}. Using a fixed value for the intrinsic luminosity as we have done in Section~\ref{sec:definition} is not ideal, even if we show in Sections~\ref{sec:int_struc_238} and \ref{sec:int_struc_1685} that using the age-luminosity fit of \cite{Mordasini2020} together with the inferred interior structure for a water-poor prior leads to intrinsic luminosities that are of the same order of magnitude as the chosen value of $10^{21}$~erg~s$^{-1}$. Furthermore, we also run our evaporation models for different values of the planetary intrinsic luminosity to mitigate this effect. The influence that the adopted value for the planetary intrinsic luminosity can have on the radii of planets with H/He dominated atmospheres is visualised in Figure~\ref{fig:triangle_luminosity}, where we show the range of possible radius values given different intrinsic luminosities. We study values between $L_{min}$\,=\,$10^{20}$~erg/s\,=\,0.25~L$_\oplus$ and $L_{max}$\,=\,$10^{22}$~erg/s\,=\,25~L$_\oplus$, both at an equilibrium temperature of 1000~K (similar to TOI-1685~b) and 1400~K (similar to TOI-238~b). Compared to mass-radius models calculated at $L$\,=\,$10^{21}$~erg/s, this can cause the upper boundary of the triangle to be up to 13\% higher assuming $L_{max}$ and up to 3.5\% lower assuming $L_{min}$ at an equilibrium temperature of 1400~K for a planet with a mass of 4~M$_\oplus$. Conversely, at 1000~K we find that the upper boundary can be up to 17\% higher and 5\% lower for a planet with a mass of 2.5~M$_\oplus$. The effect becomes smaller for higher planetary masses. While a shift of the boundary to higher values is not too concerning and simply means that our demographic analysis might not catch all potential hot water worlds, the possible shift to lower radii for planets with a lower intrinsic luminosity means that planets near the upper triangle boundary should be investigated especially carefully.

What is also not taken into account are interactions between the volatile envelope and the molten mantle layer below. Potentially, large reservoirs of dissolved hydrogen could buffer the evaporation of pure hydrogen envelopes and thereby increase the atmospheric lifetime \citep{Chachan+Stevenson2018}. Finally, in this work, we mostly consider water as a heavy volatile. However, it is of course also possible for the atmospheres of the planets in the triangles to contain other heavy volatiles, such as CO$_2$, CO, or CH$_4$. Only atmospheric follow-up observations, e.g. with JWST, can break this remaining degeneracy.

\subsection{Simulated spectra for TOI-238~b}
As our analysis in Section~\ref{sec:TOI-238b} found TOI-238~b to be a promising candidate for a hot water world, we simulated theoretical spectra for this planet to test its observability with JWST. We first calculated the atmospheric chemical composition with the open-source chemistry code \textsc{FastChem}\footnote{\url{https://github.com/NewStrangeWorlds/FastChem}} \citep{Kitzmann2024MNRAS.527.7263K, Stock2022MNRAS.517.4070S, Stock2018MNRAS.479..865S}. We considered four different scenarios: three different water mass fractions $Z_\mathrm{H2O}$ of 0.1, 0.5, and 0.95, as well as one case with scaled solar-element abundances of $Z=0.1$. To predict the chemical composition, we assumed an isothermal temperature profile with the planet's equilibrium temperature from Table~\ref{tab:planetary_params}. \textsc{FastChem} was run using the rainout condensation approach. This effectively removed all of the refractory elements from the upper atmosphere for the case with the scaled solar abundances. Condensation was also turned on for the water-rich case. However, due to the high temperatures, water did not condense in any of the considered cases.

As expected, the $Z_\mathrm{H2O}$-cases are heavily dominated by water and molecular hydrogen. Other molecules in the gas phase, such as molecular oxygen (\ce{O2}) or the hydroxide radical (\ce{OH}) are only minor constituents. The case with the scaled solar abundances, on the other hand, has a gas-phase composition that also includes other important molecules and atoms, such as carbon monoxide (\ce{CO}), carbon dioxide (\ce{CO2}), sodium (Na), or potassium (K), for example. As mentioned above, due to the use of the rainout chemistry, refractory elements, such as iron, calcium, or magnesium are condensed out in the lower atmosphere.

\begin{figure}
    \centering
    \includegraphics[width=\linewidth]{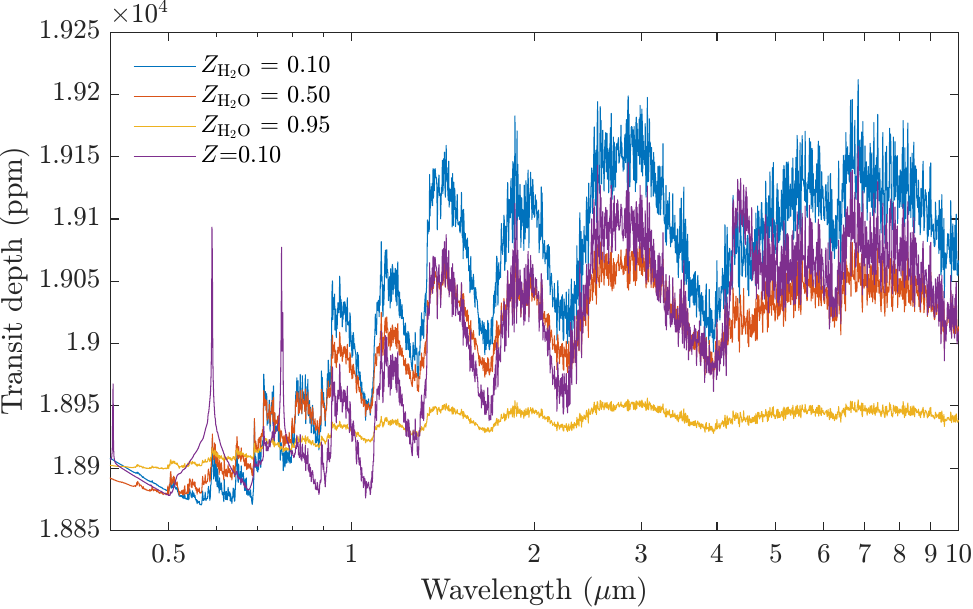}
    \caption{Theoretical spectra of TOI-238~b for three water-rich cases with different water mass fractions $Z_\mathrm{H2O}$ and one case with scaled solar-element abundances. For the latter case, $Z$ refers to the masses of the metals in the element mixture. All spectra are normalised, such that integrated over the CHEOPS bandpass they all yield the measured radius ratio from Table~\ref{tab:planetary_params}.}
    \label{fig:simulated_spectrum}
\end{figure}

For the calculation of the transmission spectra we used the open-source Bern Atmospheric Retrieval code\footnote{\textsc{BeAR} is an extended and improved version of the previous \textsc{Helios-r2} retrieval model \citep{Kitzmann2020ApJ...890..174K}. It is available on GitHub in the repository \url{https://github.com/NewStrangeWorlds/BeAR}.} (\textsc{BeAR}) in its forward-modelling mode. The transmission spectra were normalised such that integrated over the CHEOPS bandpass they yield the radius ratio from Table~\ref{tab:planetary_params}. The resulting transmission spectra are shown in Fig.~\ref{fig:simulated_spectrum}.

As the results shown in Fig. \ref{fig:simulated_spectrum} clearly suggest, the $Z_\mathrm{H2O}$-cases show strong absorption bands by \ce{H2O}, as well as a Rayleigh-scattering slope in the optical caused by \ce{H2O}, \ce{H2}, and \ce{He}. The heavily water-dominated case with $Z_\mathrm{H2O}=0.95$ exhibits variations in the transit depth of less than 20 ppm, owing to the high mean molecular weight of the atmosphere. Such an atmosphere would very likely be unobservable with the James Webb Telescope. The other two $Z_\mathrm{H2O}$-cases, on the other hand, have variations of 100 ppm or more and, thus, are potentially observable. 

The spectrum for the case with scaled solar element abundances is also heavily dominated by water absorption bands. Around 4.5 $\mu$m, a strong \ce{CO2} absorption feature is visible. Additional molecular species with important contributions to the spectrum are hydrogen sulfide (\ce{H2S}) and CO. In the optical, the strong resonance lines of Na and K are clearly visible. 

\subsection{Links to planet formation theory}
One of the most important predictions of planet formation theory is the existence of many close-in, low-mass, water-rich planets, caused by large-scale migration from outside the iceline. This goes back as far as \cite{Ward1997}, who shows that planets with masses of up to $\sim$10~M$_\oplus$ are those with the highest expected migration rates, with more modern migration models revealing similar results \citep{Tanaka+2002, Paardekooper+2011}. Identifying more planets that fall in the Hot Water World triangles could help to observationally confirm the existence of these planets.

In the following, we use a synthetic planetary population generated using the Bern model of planet formation and evolution \citep{Alibert+2005, Mordasini+2009, Emsenhuber+2021a, Emsenhuber+2021b} to illustrate the predicted water mass fractions of these close-in planets and compare to the results of our internal structure analysis above, especially for TOI-238~b, which we showed to be a promising hot water world candidate. We use the nominal population of the New Generation Planetary Population Synthesis \citep[NGPPS;][]{Emsenhuber+2021a, Emsenhuber+2021b} for solar mass stars with a longer formation phase \citep{Emsenhuber+2023} and an improved evaporation model \citep{Burn+2024}. This population was created assuming a fixed stellar mass of 1~M$_\odot$ and consists of 1000 synthetic planetary systems. 

For each synthetic planetary system, initial conditions for the protoplanetary disk were sampled using a Monte Carlo approach from observationally informed distributions. 100 randomly distributed lunar-mass planetary embryos were then placed in this protoplanetary disk, whose accretion of planetesimals and gas was then modelled during a 100~Myr formation phase, along with the dynamical interactions between the embryos and their orbital migration driven by the evolving gas disk. As a second step, the atmospheric photoevaporation of water and H/He, cooling and contraction of the planetary interior and migration due to stellar tides of each planet was modelled during a 5~Gyr evolution phase. The accreted water was assumed to be uniformly mixed with any present H/He at this stage.

\begin{figure}
    \centering
    \includegraphics[width=\linewidth]{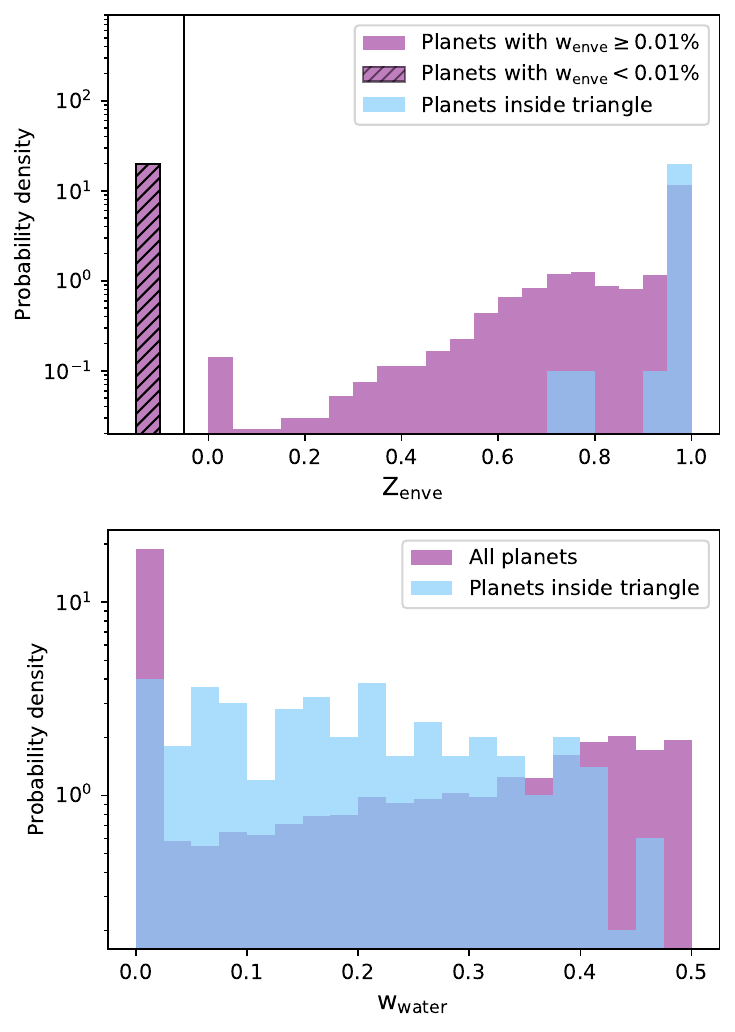}
    \caption{Histograms of the water content of the planets in the synthetic NGPPS population for solar mass stars, generated using the Bern Model of planet formation and evolution. The top panel shows the water mass fraction with respect to the mass of the volatile envelope layer of each planet, the bottom panel the water mass fraction with respect to the total planetary mass. Each panel contains two histograms, one showing all the planets in the population with masses between 0.5~and~30~M$_\oplus$ and within 2~AU (purple) and one showing the synthetic planets that fall into the Hot Water World triangles with an equilibrium temperature of at least 600~K (light blue). For the top panel, the former was split into two separate subgroups; planets with an envelope mass fraction of at least 0.01\% and planets with lower envelope mass fractions, for which defining an envelope water mass fraction is difficult (shaded in black).}
    \label{fig:Bern_model_wmf}
\end{figure}

Figure~\ref{fig:Bern_model_wmf} shows histograms of the predicted water mass fractions of the planets in this synthetic population, once with respect to the mass of the volatile envelope (top) and once with respect to the total planetary mass (bottom). This was done once for planets that lie inside the Hot Water World triangles with equilibrium temperatures of at least 600~K (light blue) and once for all planets in the population within 2~AU (purple). For both sets of planets, we filtered for planetary masses between 0.5~and~30~M$_\oplus$. 

For the histogram showing the water mass fraction with respect to the volatile envelope, the population is split into planets with an envelope mass fraction of at least 0.01\% and planets with an envelope mass fraction of less than that, as for these latter ones it is difficult to define a water mass fraction with respect to the envelope mass. When looking at the corresponding histograms, we can see that the planets inside the triangles have envelopes that are mostly made up of water, while the full planet population also includes planets with more mixed envelopes. This result is somewhat expected as the planets inside the triangles must, by definition, contain less than 1\% H/He in mass. It is also in agreement with the envelope water mass fractions we find in our internal structure analysis for TOI-238~b for a formation scenario outside the iceline (see Fig.~\ref{fig:int_struct_238}).

Pebble-based formation and evolution models predict that the total planetary water mass fractions are bimodally distributed with peaks at 0~and~50\% \citep{Venturini+2020a,Brugger+2020}, which implies that observed planets whose bulk properties only allow for water mass fractions smaller than 50\% could not have obtained this potential water through accretion of ices outside the iceline \citep[e.g.][for TOI-270~d]{Benneke+2024}. However, it is important to note that planetesimal-based models, like the one used here, do allow for intermediate water mass fractions. Indeed, the histograms of the water mass fractions with respect to the total planetary mass in the NGPPS population (bottom panel of Figure~\ref{fig:Bern_model_wmf}) show that we do not see a bimodal distribution but rather a strong peak around 0\% and then a mostly linear increase towards water mass fractions of up to 50\%. For planets that lie inside the triangles, the histograms show a distribution that is more or less uniform among the full range of water mass fractions, up to a maximum value of around 43\%.

One limitation of this model is that the geophysical evolution of each planet is not currently considered and all water is contained in the outer volatile layer, even though a large amount of water could be dissolved in the planets mantle and core \citep{Dorn+Lichtenberg2021,Vazan+2022,Luo+2024}. As discussed in \cite{Egger+2024} and \cite{Haldemann+2024}, the same is true for our internal structure model, meaning that the water mass fractions inferred by \texttt{plaNETic} are not referring to the total amount of water on the planet, but only to the water not dissolved in the interior. This complicates a comparison of the total water content between the two models. 

Furthermore, there is also the possibility of water forming endogenically through interactions between a molten magma ocean and hydrogen \citep{Kite+2020}. This makes it more difficult to interpret close-in planets with observationally detected water, which could either have formed outside the iceline and migrated inwards, thereby observationally confirming large-scale migration, or could have formed more or less in-situ with water forming endogenically.

\section{Summary}
\label{sec:conclusions}
In this paper, we introduced the concept of Hot Water World triangles, which we define as triangular regions in the mass-radius space where a close-in planet can only lie if its volatile layer consists, at least to some part, of volatiles heavier than H/He. This is because at close orbital separations, a low-mass H/He envelope would be evaporated quickly due to high levels of high-energy irradiation the planet receives from its host star, whereas heavier volatiles, such as water, have a higher mean molecular weight and are therefore less affected. 

We used this concept to identify promising candidates for close-in water world planets, simply based on precise measurements of their orbital periods, radii, and masses. Such planets represent interesting candidates for further atmospheric characterisation, e.g. with the James Webb Space Telescope. Our demographic study of known exoplanets with equilibrium temperatures of at least 600~K and with precisely known mass and radius values showed that planets that fall into these regions are rare, with only ten well-characterised planets that lie fully within the Hot Water World triangles at the 2$\sigma$ level: HD~86226~c \citep{Teske+2020}, Kepler-23~b \citep{Leleu+2023}, Kepler-60~b and~c \citep{Leleu+2023}, Kepler-68~b \citep{Bonomo+2023}, TOI-1685~b \citep{Luque+Palle2022}, TOI-544~b \citep{Osborne+2024}, TOI-733~b \citep{Georgieva+2023}, Wolf~503~b \citep{Polanski+2021}, and $\pi$~Men~c \citep{Damasso+2020}. We further observed that especially for planets coinciding with these triangular regions, there are cases of planets where contradicting mass and/or radius values exist in the literature, making it unclear whether these planets actually lie within the Hot Water World triangles.

For this reason, we use CHEOPS observations to refine the radii of some of these planets, as well as following up on potential new hot water world candidates. In this paper, we presented the collected CHEOPS data for the first two targets of our GTO programme `Hot Water Worlds', TOI-238~b and TOI-1685~b, which we fitted along with all available TESS sectors. Our derived radius for TOI-238~b, $1.559\pm0.047$~R$_\oplus$, is $\sim$1.8$\sigma$ ($\sim$3.3$\sigma$ using our own smaller uncertainties) larger than the value reported in the discovery paper by \cite{Mascareno+2024}, but agrees well with the value reported by \cite{Mistry+2024}, who first validated the planet. With a mass of $3.37\pm0.49$~M$_\oplus$, we therefore found that this planet does in fact lie inside the corresponding Hot Water World triangle within 1$\sigma$ of both mass and radius. Conversely, for TOI-1685~b, we find a radius of $1.421\pm0.060$~R$_\oplus$, which is in agreement with the values derived by \cite{Hirano+2021} and \cite{Burt+2024}, but significantly smaller than the values reported by \cite{Bluhm+2021} and \cite{Luque+Palle2022}. Combined with a mass of $3.07^{+0.34}_{-0.33}$~M$_\oplus$, this leads to the conclusion that the planet lies, at the 1$\sigma$ level, below the corresponding triangle. This does not mean that TOI-1685~b cannot be a hot water world, but that its interior composition is fully degenerate and it is also possible that the planet is a bare core.

We then ran an interior structure analysis for both planets using the \texttt{plaNETic} framework \citep{Egger+2024}, which for both planets showed that the derived planetary parameters are consistent with water-rich envelopes of a few percent in planetary mass, while the models for water-poor envelopes tend towards envelope mass fractions of the order of $10^{-6}$, which is the lower bound of the chosen prior. These very low-mass volatile layers are made up of almost pure H/He. While the planetary parameters for TOI-1685~b are also compatible with a bare core, this is not the case for TOI-238~b at a 1$\sigma$ level.

Next, we studied the potential lifetime of these H/He and water envelopes for both planets. For pure H/He, we find very short lifetimes of around 0.4-1.3~Myr on TOI-238~b and of only $\sim$0.1~Myr on TOI-1685~b, meaning that a H/He atmosphere would not be stable on either planet. At the same time, we find that an H$_2$O atmosphere could have survived on both planets, depending on whether the corresponding host stars evolved as fast or slow rotators and the assumed heating efficiency parameters. Especially if adopting a heating efficiency parameter $\eta$\,=\,0.05 as favoured by our hydrodynamic models, we find that H$_2$O atmospheres remain stable across most other parameter combinations considered.

As we found TOI-238~b to be a promising hot water world candidate, we simulated theoretical spectra assuming different atmospheric water mass fractions, thereby investigating the potential observability of the spectral features with JWST. Finally, we discussed links to planet formation theory, as the detection of a large number of close-in water-rich planets would confirm one of its main predictions, large-scale migration of small planets inwards from beyond the iceline. We found that for a synthetic population of planets, the total water mass fractions of planets in the triangles are distributed more or less uniformly from a few percent up to a maximum value of around 45\%, while the distribution in the full population showcases a strong peak at 0\% and then a linear increase towards 50\% water. However, the origin of the water in the detected planets is not clear, as it could also have formed endogenically through interactions of a H/He atmosphere with a molten magma ocean.

\begin{acknowledgements}
We thank an anonymous referee for valuable comments
that helped improve the manuscript. CHEOPS is an ESA mission in partnership with Switzerland with important contributions to the payload and the ground segment from Austria, Belgium, France, Germany, Hungary, Italy, Portugal, Spain, Sweden, and the United Kingdom. The CHEOPS Consortium would like to gratefully acknowledge the support received by all the agencies, offices, universities, and industries involved. Their flexibility and willingness to explore new approaches were essential to the success of this mission. CHEOPS data analysed in this article will be made available in the CHEOPS mission archive (\url{https://cheops.unige.ch/archive_browser/}). 
This work has been carried out within the framework of the NCCR PlanetS supported by the Swiss National Science Foundation under grants 51NF40\_182901 and 51NF40\_205606. 
YAl acknowledges support from the Swiss National Science Foundation (SNSF) under grant 200020\_192038. 
TWi acknowledges support from the UKSA and the University of Warwick. 
ABr was supported by the SNSA. 
MNG is the ESA CHEOPS Project Scientist and Mission Representative. BMM is the ESA CHEOPS Project Scientist. KGI was the ESA CHEOPS Project Scientist until the end of December 2022 and Mission Representative until the end of January 2023. All of them are/were responsible for the Guest Observers (GO) Programme. None of them relay/relayed proprietary information between the GO and Guaranteed Time Observation (GTO) Programmes, nor do/did they decide on the definition and target selection of the GTO Programme.
ML acknowledges support of the Swiss National Science Foundation under grant number PCEFP2\_194576. 
S.G.S. acknowledge support from FCT through FCT contract nr. CEECIND/00826/2018 and POPH/FSE (EC). 
The Portuguese team thanks the Portuguese Space Agency for the provision of financial support in the framework of the PRODEX Programme of the European Space Agency (ESA) under contract number 4000142255. 
MF and CMP gratefully acknowledge the support of the Swedish National Space Agency (DNR 65/19, 174/18). 
DG gratefully acknowledges financial support from the CRT foundation under Grant No. 2018.2323 “Gaseousor rocky? Unveiling the nature of small worlds”. 
We acknowledge financial support from the Agencia Estatal de Investigación of the Ministerio de Ciencia e Innovación MCIN/AEI/10.13039/501100011033 and the ERDF “A way of making Europe” through projects PID2019-107061GB-C61, PID2019-107061GB-C66, PID2021-125627OB-C31, and PID2021-125627OB-C32, from the Centre of Excellence “Severo Ochoa” award to the Instituto de Astrofísica de Canarias (CEX2019-000920-S), from the Centre of Excellence “María de Maeztu” award to the Institut de Ciències de l’Espai (CEX2020-001058-M), and from the Generalitat de Catalunya/CERCA programme. 
DBa, EPa, and IRi acknowledge financial support from the Agencia Estatal de Investigación of the Ministerio de Ciencia e Innovación MCIN/AEI/10.13039/501100011033 and the ERDF “A way of making Europe” through projects PID2019-107061GB-C61, PID2019-107061GB-C66, PID2021-125627OB-C31, and PID2021-125627OB-C32, from the Centre of Excellence “Severo Ochoa'' award to the Instituto de Astrofísica de Canarias (CEX2019-000920-S), from the Centre of Excellence “María de Maeztu” award to the Institut de Ciències de l’Espai (CEX2020-001058-M), and from the Generalitat de Catalunya/CERCA programme. 
SCCB acknowledges the support from Fundação para a Ciência e Tecnologia (FCT) in the form of work contract through the Scientific Employment Incentive program with reference 2023.06687.CEECIND. 
LBo, VNa, IPa, GPi, RRa, and GSc acknowledge support from CHEOPS ASI-INAF agreement n. 2019-29-HH.0. 
CBr and ASi acknowledge support from the Swiss Space Office through the ESA PRODEX program. 
ACC acknowledges support from STFC consolidated grant number ST/V000861/1, and UKSA grant number ST/X002217/1. 
ACMC acknowledges support from the FCT, Portugal, through the CFisUC projects UIDB/04564/2020 and UIDP/04564/2020, with DOI identifiers 10.54499/UIDB/04564/2020 and 10.54499/UIDP/04564/2020, respectively. 
A.C., A.D., B.E., K.G., and J.K. acknowledge their role as ESA-appointed CHEOPS Science Team Members. 
P.E.C. is funded by the Austrian Science Fund (FWF) Erwin Schroedinger Fellowship, program J4595-N. 
This project was supported by the CNES. 
A.De. 
This work was supported by FCT - Funda\c{c}\~{a}o para a Ci\^{e}ncia e a Tecnologia through national funds and by FEDER through COMPETE2020 through the research grants UIDB/04434/2020, UIDP/04434/2020, 2022.06962.PTDC. 
O.D.S.D. is supported in the form of work contract (DL 57/2016/CP1364/CT0004) funded by national funds through FCT. 
B.-O. D. acknowledges support from the Swiss State Secretariat for Education, Research and Innovation (SERI) under contract number MB22.00046. 
ADe, BEd, KGa, and JKo acknowledge their role as ESA-appointed CHEOPS Science Team Members. 
This project has received funding from the Swiss National Science Foundation for project 200021\_200726. It has also been carried out within the framework of the National Centre of Competence in Research PlanetS supported by the Swiss National Science Foundation under grant 51NF40\_205606. The authors acknowledge the financial support of the SNSF. 
M.G. is an F.R.S.-FNRS Senior Research Associate. 
CHe acknowledges the European Union H2020-MSCA-ITN-2019 under GrantAgreement no. 860470 (CHAMELEON), and the HPC facilities at the Vienna Science Cluster (VSC). 
K.W.F.L. was supported by Deutsche Forschungsgemeinschaft grants RA714/14-1 within the DFG Schwerpunkt SPP 1992, Exploring the Diversity of Extrasolar Planets. 
This work was granted access to the HPC resources of MesoPSL financed by the Region Ile de France and the project Equip@Meso (reference ANR-10-EQPX-29-01) of the programme Investissements d'Avenir supervised by the Agence Nationale pour la Recherche. 
Support for this work was provided by NASA through the NASA Hubble Fellowship grant \#HST-HF2-51559.001-A awarded by the Space Telescope Science Institute which is operated by the Association of Universities for Research in Astronomy Inc. for NASA under contract NAS5-26555. 
PM acknowledges support from STFC research grant number ST/R000638/1. 
This work was also partially supported by a grant from the Simons Foundation (PI Queloz, grant number 327127). 
NCSa acknowledges funding by the European Union (ERC, FIERCE, 101052347). Views and opinions expressed are however those of the author(s) only and do not necessarily reflect those of the European Union or the European Research Council. Neither the European Union nor the granting authority can be held responsible for them. 
GyMSz acknowledges the support of the Hungarian National Research, Development and Innovation Office (NKFIH) grant K-125015, a a PRODEX Experiment Agreement No. 4000137122, the Lend\"ulet LP2018-7/2021 grant of the Hungarian Academy of Science and the support of the city of Szombathely. 
V.V.G. is an F.R.S-FNRS Research Associate. 
JV acknowledges support from the Swiss National Science Foundation (SNSF) under grant PZ00P2\_208945. 
EV acknowledges support from the ‘DISCOBOLO’ project funded by the Spanish Ministerio de Ciencia, Innovación y Universidades undergrant PID2021-127289NB-I00. 
NAW acknowledges UKSA grant ST/R004838/1. 
\\ \textit{Software.} The following software packages have been used for this publication: Python-numpy \citep{numpy}, Python-pandas \citep{pandas}, Python-matplotlib \citep{matplotlib}, Python-seaborn \citep{seaborn}, Python-tensorflow \citep{tensorflow}, Python-scipy \citep{scipy}.
\end{acknowledgements}

%
%

\bibliographystyle{aa} 
\bibliography{biblio} 

\onecolumn
\appendix

\section{Observations and data analysis}
\begin{table}[H]
\renewcommand{\arraystretch}{1.4}
    \caption{CHEOPS observation logs for TOI-1685 and TOI-238.}
    \centering
    \begin{tabular}{ccccccc}
    \toprule
    \toprule
    ID & Start Date [UTC] & Dur [orbits] & File Key & Av. eff. [\%] & RMS [ppm] & Planet\\
    \hline
    1 & 2023-07-13T16:40:20.014 & $ 7.03 $ & PR140068\_TG002301\_V0300 & $ 58 $ & $ 485 $ & TOI-238~b\\
    2 & 2023-08-15T23:34:58.185 & $ 4.02 $ & PR140068\_TG002401\_V0300 & $ 88 $ & $ 494 $ & TOI-238~b\\
    3 & 2023-08-23T15:15:35.935 & $ 4.02 $ & PR140068\_TG002402\_V0300 & $ 90 $ & $ 571 $ & TOI-238~b\\
    4 & 2023-08-28T17:50:22.475 & $ 4.90 $ & PR140068\_TG002403\_V0300 & $ 86 $ & $ 469 $ & TOI-238~b\\
    5 & 2023-09-23T05:36:03.064 & $ 4.07 $ & PR140068\_TG002404\_V0300 & $ 85 $ & $ 511 $ & TOI-238~b\\
    \hline
    6 & 2023-10-20T13:09:37.032 & $ 3.53 $ & PR140068\_TG002701\_V0300 & $ 55 $ & $ 1444 $ & TOI-1685~b\\
    7 & 2023-11-05T14:46:56.459 & $ 3.50 $ & PR140068\_TG002702\_V0300 & $ 57 $ & $ 1433 $ & TOI-1685~b\\
    8 & 2023-11-10T23:18:16.432 & $ 3.56 $ & PR140068\_TG002703\_V0300 & $ 58 $ & $ 1334 $ & TOI-1685~b\\
    9 & 2023-12-03T01:43:58.488 & $ 3.55 $ & PR140068\_TG002901\_V0300 & $ 58 $ & $ 1272 $ & TOI-1685~b\\
    10 & 2023-12-06T09:31:59.965 & $ 3.92 $ & PR140068\_TG002902\_V0300 & $ 56 $ & $ 1152 $ & TOI-1685~b\\
    11 & 2023-12-09T17:30:58.821 & $ 4.13 $ & PR140068\_TG002903\_V0300 & $ 51 $ & $ 1158 $ & TOI-1685~b\\
    12 & 2023-12-14T09:17:53.524 & $ 6.42 $ & PR140068\_TG003001\_V0300 & $ 57 $ & $ 1138 $ & TOI-1685~b\\
    13 & 2023-12-22T11:19:39.043 & $ 4.48 $ & PR140068\_TG003102\_V0300 & $ 58 $ & $ 1149 $ & TOI-1685~b\\
    14 & 2023-12-15T01:40:53.155 & $ 4.54 $ & PR140068\_TG003101\_V0300 & $ 58 $ & $ 1187 $ & TOI-1685~b\\
    15 & 2023-12-19T16:45:46.147 & $ 5.01 $ & PR140068\_TG003002\_V0300 & $ 51 $ & $ 1119 $ & TOI-1685~b\\
    16 & 2023-12-20T08:48:44.599 & $ 7.14 $ & PR140068\_TG003003\_V0300 & $ 56 $ & $ 1155 $ & TOI-1685~b\\
    17 & 2023-12-28T09:31:23.599 & $ 5.03 $ & PR140068\_TG003004\_V0300 & $ 57 $ & $ 1134 $ & TOI-1685~b\\
    18 & 2024-01-07T12:16:45.308 & $ 4.92 $ & PR140068\_TG003201\_V0300 & $ 54 $ & $ 1218 $ & TOI-1685~b\\
    19 & 2024-01-09T10:59:36.437 & $ 4.92 $ & PR140068\_TG003202\_V0300 & $ 55 $ & $ 1101 $ & TOI-1685~b\\
    20 & 2024-01-12T02:50:23.636 & $ 6.13 $ & PR140068\_TG003203\_V0300 & $ 55 $ & $ 1150 $ & TOI-1685~b\\
    21 & 2024-01-12T18:53:20.312 & $ 4.92 $ & PR140068\_TG003204\_V0300 & $ 48 $ & $ 1160 $ & TOI-1685~b\\
    \bottomrule
    \end{tabular}
    \label{tab:file_keys}
    \tablefoot{One CHEOPS orbit is 98.77~minutes.}
\end{table}
\renewcommand{\arraystretch}{1}

\begin{figure}[H]
    \centering
    \includegraphics[width=0.95\textwidth]{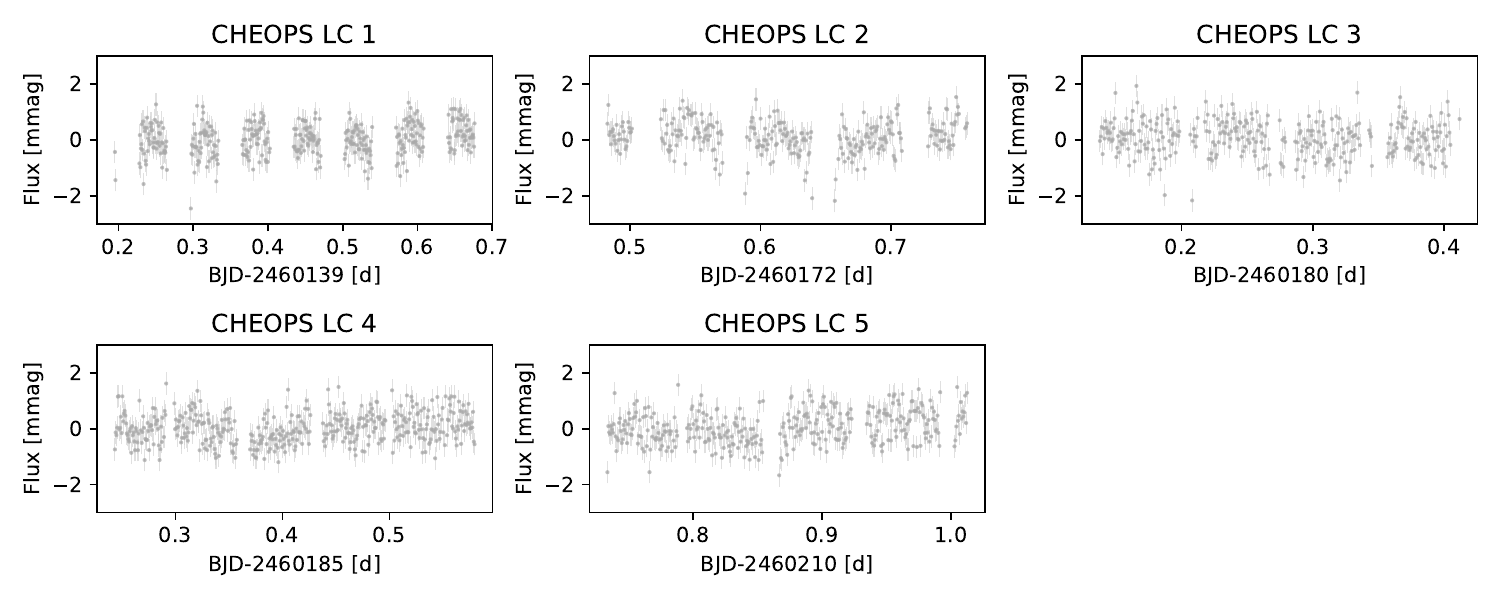}
    \caption{Undetrended CHEOPS light curves for TOI-238~b, identified by the IDs introduced in Table \ref{tab:file_keys}.}
    \label{fig:undetrended_CHEOPS_238}
\end{figure}

\begin{figure}[H]
    \centering
    \includegraphics[width=0.95\textwidth]{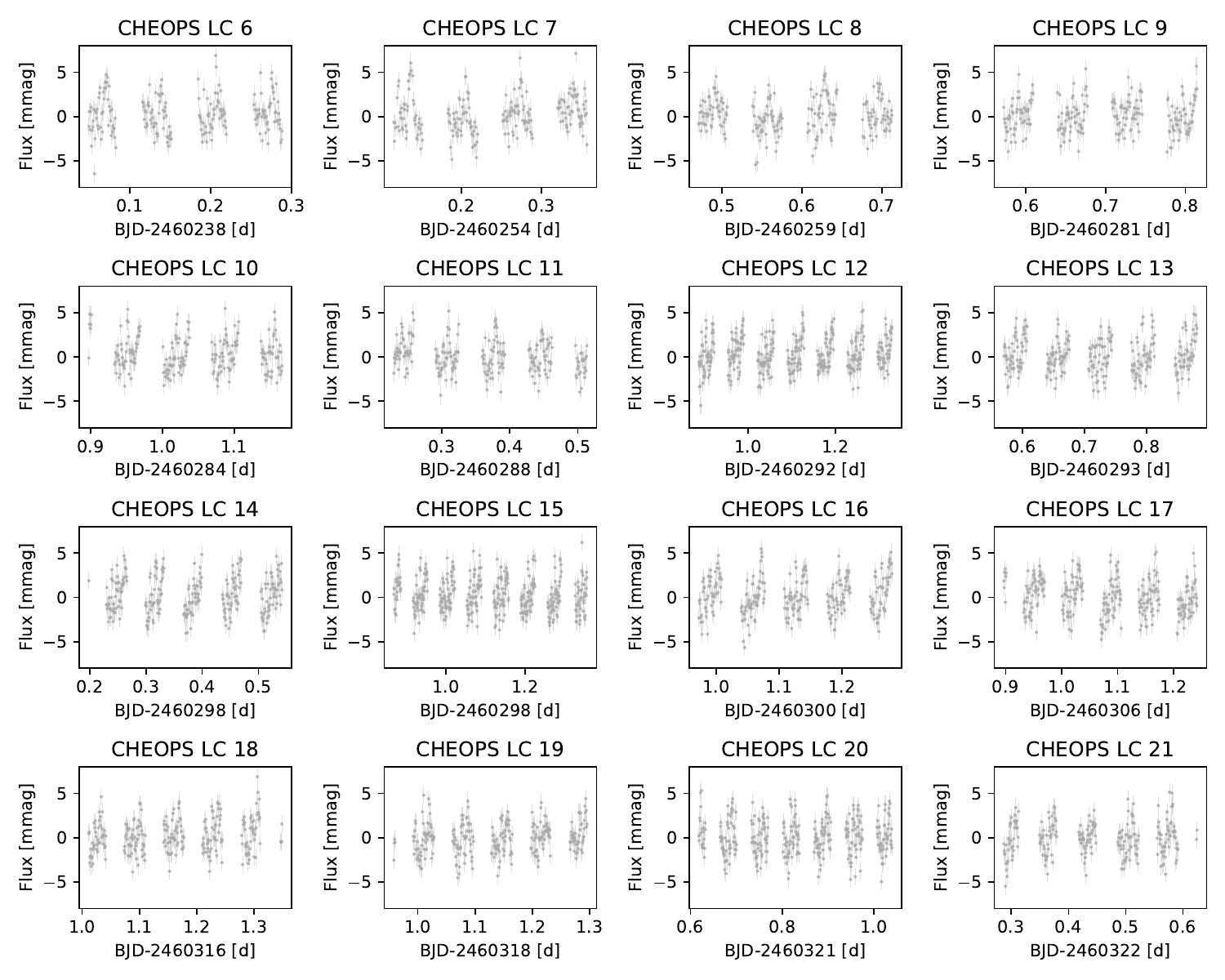}
    \caption{Undetrended CHEOPS light curves for TOI-1685~b, identified by the IDs introduced in Table \ref{tab:file_keys}.}
    \label{fig:undetrended_CHEOPS_1685}
\end{figure}

\newpage
\section{Internal structure modelling}

\begin{table}[H]
\renewcommand{\arraystretch}{1.5}
\caption{Medians and one-sigma errors for the \texttt{plaNETic} internal structure modelling posteriors for TOI-238~b.}
\centering
\begin{tabular}{r|ccc|ccc}
\hline \hline
Water prior &              \multicolumn{3}{c|}{Water-rich (e.g. formation outside iceline)} & \multicolumn{3}{c}{Water-poor (e.g. formation inside iceline)} \\
Si/Mg/Fe prior &           Stellar (A1) &       Iron-enriched (A2) &      Free (A3) &
                           Stellar (B1) &       Iron-enriched (B2) &      Free (B3) \\
\hline
w$_\textrm{core}$ [\%] &        $17_{-11}^{+11}$ &    $26_{-18}^{+19}$ &    $26_{-18}^{+21}$ &
                           $21_{-12}^{+10}$ &    $34_{-22}^{+19}$ &    $37_{-24}^{+21}$ \\
w$_\textrm{mantle}$ [\%] &      $80_{-11}^{+12}$ &    $69_{-20}^{+18}$ &    $69_{-22}^{+18}$ &
                           $78_{-10}^{+12}$ &    $66_{-19}^{+22}$ &    $63_{-21}^{+24}$ \\
w$_\textrm{envelope}$ [\%] &    $2.3_{-1.6}^{+2.9}$ &    $4.9_{-3.1}^{+4.4}$ &    $5.1_{-3.4}^{+4.9}$ &
                                $\left(1.3_{-0.1}^{+0.2}\right)$ $10^{-4}$ &    $\left(1.6_{-0.5}^{+1.2}\right)$ $10^{-4}$ &    $\left(1.9_{-0.7}^{+1.9}\right)$ $10^{-4}$ \\
\hline
Z$_\textrm{envelope}$ [\%] &        $99.9_{-1.1}^{+0.1}$ &    $99.9_{-1.2}^{+0.1}$ &    $99.9_{-1.2}^{+0.1}$ &
                           $0.5_{-0.2}^{+0.3}$ &    $0.5_{-0.2}^{+0.2}$ &    $0.5_{-0.2}^{+0.2}$ \\
\hline
x$_\textrm{Fe,core}$ [\%] &     $90.3_{-6.4}^{+6.6}$ &    $90.3_{-6.4}^{+6.6}$ &    $90.4_{-6.4}^{+6.5}$ &
                           $89.8_{-6.2}^{+7.0}$ &    $90.4_{-6.4}^{+6.5}$ &    $90.4_{-6.4}^{+6.5}$ \\
x$_\textrm{S,core}$ [\%] &      $9.7_{-6.6}^{+6.4}$ &    $9.7_{-6.6}^{+6.4}$ &    $9.6_{-6.5}^{+6.4}$ &
                           $10.2_{-7.0}^{+6.2}$ &    $9.6_{-6.5}^{+6.4}$ &    $9.6_{-6.5}^{+6.4}$ \\
\hline
x$_\textrm{Si,mantle}$ [\%] &   $41_{-7}^{+8}$ &    $34_{-9}^{+11}$ &    $27_{-19}^{+27}$ &
                           $42_{-8}^{+8}$ &    $32_{-9}^{+12}$ &    $22_{-15}^{+25}$ \\
x$_\textrm{Mg,mantle}$ [\%] &   $42_{-7}^{+7}$ &    $33_{-9}^{+11}$ &    $34_{-21}^{+25}$ &
                           $42_{-7}^{+8}$ &    $32_{-9}^{+11}$ &    $34_{-20}^{+25}$ \\
x$_\textrm{Fe,mantle}$ [\%] &   $17_{-11}^{+10}$ &    $32_{-20}^{+18}$ &    $33_{-21}^{+22}$ &
                           $15_{-9}^{+10}$ &    $36_{-22}^{+18}$ &    $39_{-24}^{+21}$ \\
\hline
\end{tabular}
\label{tab:internal_structure_results_238}
\end{table}
\renewcommand{\arraystretch}{1.0}

\begin{table}[H]
\renewcommand{\arraystretch}{1.5}
\caption{Medians and one-sigma errors for the \texttt{plaNETic} internal structure modelling posteriors for TOI-1685~b.}
\centering
\begin{tabular}{r|ccc|ccc}
\hline \hline
Water prior &              \multicolumn{3}{c|}{Formation outside iceline (water-rich)} & \multicolumn{3}{c}{Formation inside iceline (water-poor)} \\
Si/Mg/Fe prior &           Stellar (A1) &       Iron-enriched (A2) &      Free (A3) &
                           Stellar (B1) &       Iron-enriched (B2) &      Free (B3) \\
\hline
w$_\textrm{core}$ [\%] &        $21_{-14}^{+16}$ &    $32_{-21}^{+23}$ &    $29_{-20}^{+22}$ &
                           $31_{-20}^{+20}$ &    $39_{-26}^{+23}$ &    $39_{-25}^{+21}$ \\
w$_\textrm{mantle}$ [\%] &      $77_{-16}^{+14}$ &    $65_{-24}^{+22}$ &    $67_{-23}^{+20}$ &
                           $69_{-20}^{+20}$ &    $60_{-23}^{+26}$ &    $61_{-21}^{+25}$ \\
w$_\textrm{envelope}$ [\%] &    $1.2_{-0.9}^{+2.3}$ &    $3.1_{-2.2}^{+3.8}$ &    $2.5_{-1.8}^{+3.3}$ &
                           $\left(1.5_{-0.4}^{+1.6}\right)$ $10^{-4}$ &    $\left(2.2_{-0.9}^{+3.4}\right)$ $10^{-4}$ &    $\left(1.9_{-0.7}^{+2.2}\right)$ $10^{-4}$ \\
\hline
Z$_\textrm{envelope}$ [\%] &        $99.8_{-1.2}^{+0.1}$ &    $99.9_{-1.3}^{+0.1}$ &    $99.9_{-1.3}^{+0.1}$ &
                           $0.5_{-0.2}^{+0.2}$ &    $0.5_{-0.2}^{+0.2}$ &    $0.5_{-0.2}^{+0.2}$ \\
\hline
x$_\textrm{Fe,core}$ [\%] &     $90.3_{-6.4}^{+6.6}$ &    $90.4_{-6.4}^{+6.5}$ &    $90.4_{-6.4}^{+6.5}$ &
                           $90.3_{-6.4}^{+6.6}$ &    $90.3_{-6.4}^{+6.5}$ &    $90.3_{-6.4}^{+6.6}$ \\
x$_\textrm{S,core}$ [\%] &      $9.7_{-6.6}^{+6.4}$ &    $9.6_{-6.5}^{+6.4}$ &    $9.6_{-6.5}^{+6.4}$ &
                           $9.7_{-6.6}^{+6.4}$ &    $9.7_{-6.5}^{+6.4}$ &    $9.7_{-6.6}^{+6.4}$ \\
\hline
x$_\textrm{Si,mantle}$ [\%] &   $34_{-16}^{+18}$ &    $24_{-13}^{+17}$ &    $24_{-16}^{+25}$ &
                           $25_{-16}^{+24}$ &    $22_{-13}^{+18}$ &    $21_{-15}^{+24}$ \\
x$_\textrm{Mg,mantle}$ [\%] &   $43_{-18}^{+16}$ &    $29_{-15}^{+18}$ &    $32_{-19}^{+23}$ &
                           $42_{-23}^{+22}$ &    $28_{-15}^{+20}$ &    $34_{-19}^{+25}$ \\
x$_\textrm{Fe,mantle}$ [\%] &   $21_{-13}^{+17}$ &    $44_{-26}^{+22}$ &    $39_{-23}^{+20}$ &
                           $27_{-18}^{+23}$ &    $46_{-28}^{+22}$ &    $39_{-25}^{+21}$ \\
\hline
\end{tabular}
\label{tab:internal_structure_results_1685}
\end{table}
\renewcommand{\arraystretch}{1.0}

\end{document}